\documentclass[a4paper,fleqn,usenatbib]{mnras}

\usepackage{newtxtext,newtxmath}

\usepackage[T1]{fontenc}
\usepackage{ae,aecompl}

\usepackage{graphicx}   % Including figure files
\usepackage{amsmath}    % Advanced maths commands
\usepackage{amssymb}    % Extra maths symbols
\usepackage{longtable} % Allows table to span multiple pages
\def\aj{AJ}% % Astronomical Journal
\def\actaa{Acta Astron.}% % Acta Astronomica
\def\apj{ApJ}% % Astrophysical Journal
\def\apjl{ApJ}% % Astrophysical Journal, Letters
\def\apjs{ApJS}% % Astrophysical Journal, Supplement
\def\ao{Appl.~Opt.}% % Applied Optics
\def\mnras{MNRAS}% % Monthly Notices of the RAS
\def\pasp{PASP}% % Publications of the ASP
\def\pasj{PASJ}% % Publications of the ASJ
\def\rmxaa{Rev. Mexicana Astron. Astrofis.}% % Revista Mexicana de Astronomia
\def\solphys{Sol.~Phys.}% % Solar Physics
\title[Astrometric and photometric study of NGC 6067, NGC 2506 and IC 4651]
      {Astrometric and photometric study of NGC 6067, NGC 2506 and IC 4651 open clusters based on wide-field ground and Gaia DR2 data}

\author[Rangwal et al.]
   {Geeta Rangwal$^{1}$\thanks{E-mail: geetarangwal91@gmail.com}, 
    R. K. S. Yadav$^{2}$, Alok Durgapal$^{1}$, D. Bisht$^{3}$,
    D. Nardiello$^{4,5}$\\
    \\ 
    $^{1}$ Center of Advanced Study, Department of Physics,
           D. S. B. Campus, Kumaun University Nainital 263002, India\\
    $^{2}$ Aryabhatta Research Institute of Observational Sciences, 
           Manora Peak, Nainital 263129, India\\
    $^{3}$ Key Laboratory for Researches in Galaxies and Cosmology, 
           University of Science and Technology of China,   \\
           Chinese Academy of Sciences, Hefei, Anhui, 230026, China  \\
    $^{4}$Dipartimento di Fisica e Astronomia ``Galileo Galilei'', Universit\`a
di Padova, Vicolo dell'Osservatorio 3, Padova IT-35122 \\
    $^{5}$Istituto Nazionale di Astrofisica - Osservatorio Astronomico di
Padova, Vicolo dell'Osservatorio 5, Padova, IT-35122 \\
    }

\date{Accepted XXX. Received YYY; in original form ZZZ}

\pubyear{2019}

\begin{document}
\label{firstpage}
\pagerange{\pageref{firstpage}--\pageref{lastpage}}
\maketitle

% Abstract of the paper

\begin{abstract}

We present an analysis of three southern open star clusters NGC 6067,
NGC 2506 and IC 4651 using wide-field photometric and 
Gaia DR2 astrometric data. They are poorly studied clusters. 
We took advantage of the synergy between Gaia DR2 high precision astrometric
measurements and ground based wide-field photometry to
isolate cluster members and further study these clusters. 
We identify the cluster members using proper
motions, parallax and colour-magnitude diagrams. Mean proper motion of the clusters
in $\mu_{\alpha}cos\delta$ and $\mu_\delta$ is estimated as $-1.90\pm0.01$ and
$-2.57\pm0.01$ mas yr$^{-1}$ for NGC 6067, $-2.57\pm0.01$ and $3.92\pm0.01$
mas yr$^{-1}$ for NGC 2506 and $-2.41\pm0.01$ and $-5.05\pm0.02$ mas yr$^{-1}$
for IC 4651. Distances are estimated as $3.01\pm0.87$, $3.88\pm0.42$ and
$1.00\pm0.08$ kpc for the clusters NGC 6067, NGC 2506 and IC 4651 respectively using parallaxes
taken from Gaia DR2 catalogue. Galactic orbits are determined for these clusters using
Galactic potential models. We find that these clusters have circular
orbits. Cluster radii
are determined as 10$^\prime$ for NGC 6067, 12$^\prime$
for NGC 2506 and 11$^\prime$ for IC 4651. Ages of the 
clusters estimated by isochrones fitting are 
$66 \pm 8$ Myr, $2.09 \pm 0.14$ Gyr and
$1.59 \pm 0.14$ Gyr for NGC 6067, NGC 2506 and IC 4651 respectively.
Mass function slope for the entire region of cluster NGC 2506
is found to be comparable with the Salpeter value in the mass range 0.77 - 1.54 $M_{\odot}$.
The mass function analysis
shows that the slope becomes flat when one goes from halo to core region in
all the three clusters. A comparison of dynamical age with cluster's age indicates
that NGC 2506 and IC 4651 are dynamically relaxed clusters.

\end{abstract}


%%%%%%%%%%%%%%%%% BODY OF PAPER %%%%%%%%%%%%%%%%%%

% sec:intro
%________________________________________________________________

\section{Introduction} \label{sec:intro}

\begin{figure*}
    \centering
    \includegraphics[width=8cm, height=8cm]{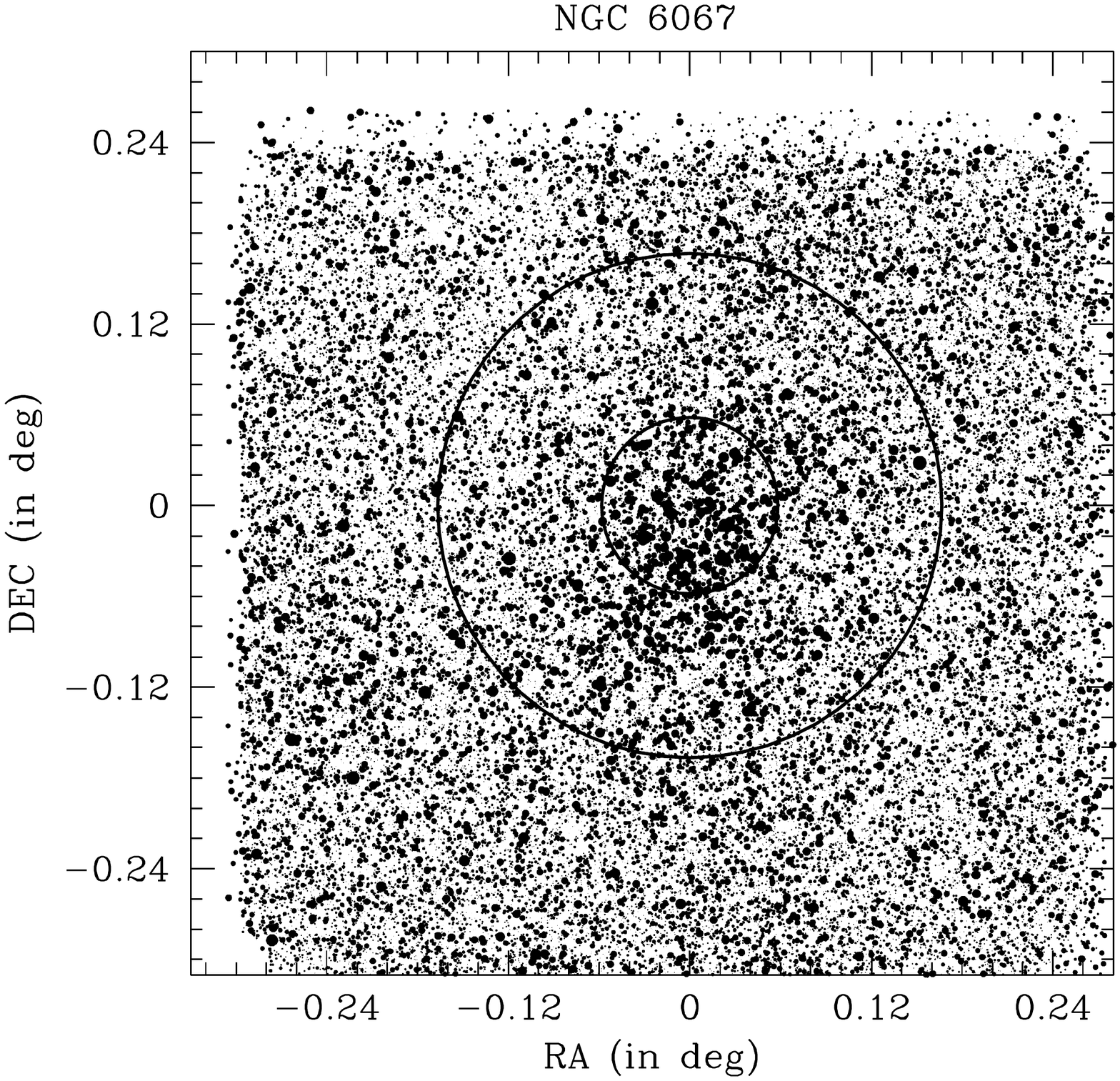}
    \includegraphics[width=8cm, height=8cm]{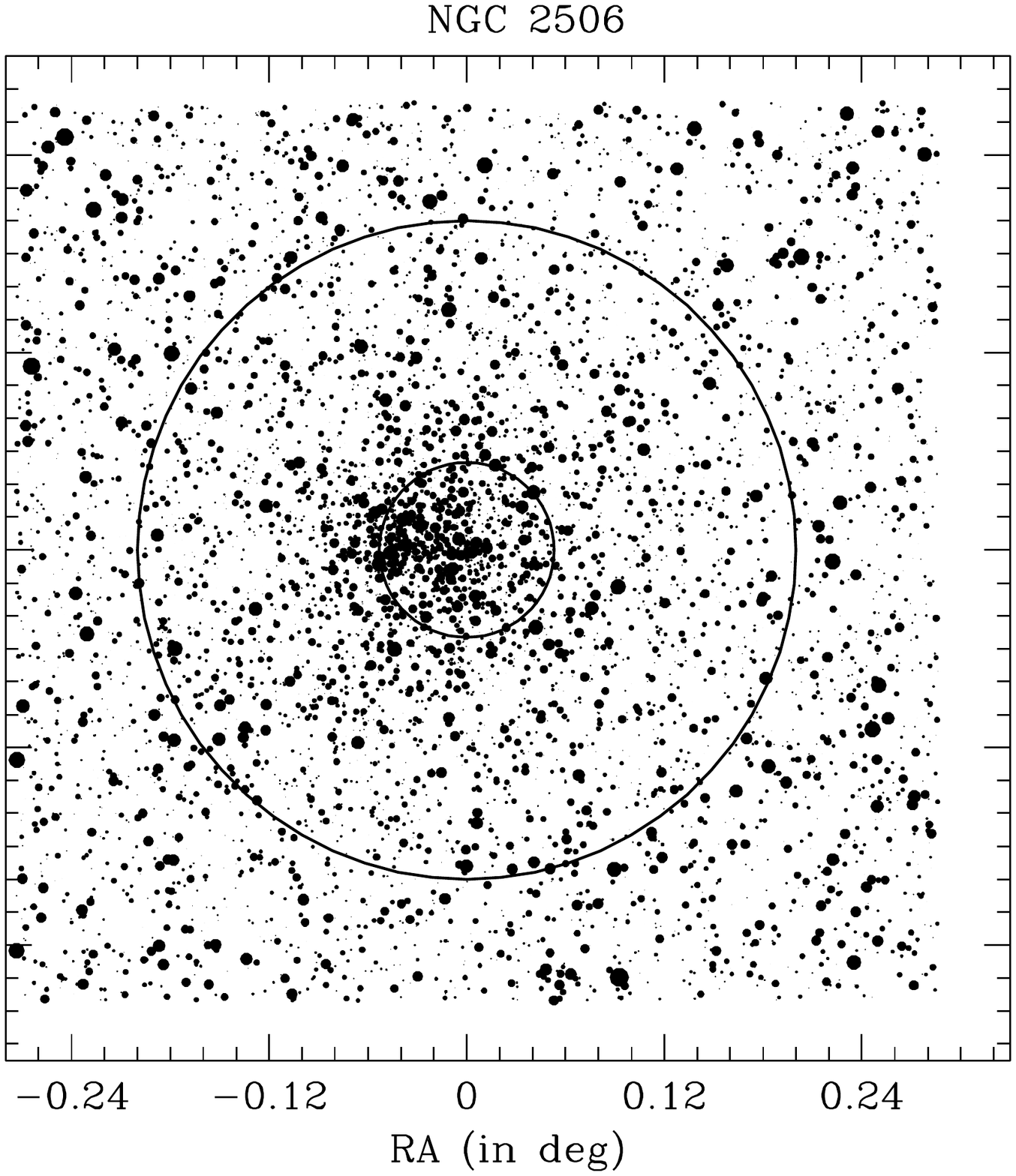}
    \caption{ Identification maps
    for the clusters
       NGC 6067 and NGC 2506. Right ascension (RA) and declination (DEC) of
     the stars are in
     degrees. Filled circles
     of different sizes represents the brightness of the stars. The
    smallest size of filled circle denotes stars of $V$ $\sim 18$ mag.}
    \label{id}
  \end{figure*}

Open star clusters are found in the plane of the Galaxy and span a 
large range in age and distance. It is easier to calculate parameters
of stars in a cluster than field stars.
All stars in a cluster are formed by the same molecular cloud and because 
of this they are dynamically associated with each other and have
nearly same distance from us. This makes them an
ideal object to study the evolution and dynamics of 
the Galactic disc. For such studies, it is very important to know
their true members and basic parameters such as age, distance, metallicity, and
reddening.

Since open clusters are usually projected against the Galactic disc stars, it
is very difficult to isolate cluster members from field stars in the 
absence of radial velocity and proper motion data. 
Accurate proper motion data are required to 
eliminate field stars. The proper motion data provided
by Gaia mission are very valuable to separate the cluster
members from field stars.
It is also expected that the Gaia mission
will completely transform our knowledge of
the structure and dynamics of the Galaxy \citep{2016A&A...595A...1G}.
The 2$^{nd}$ data release (DR2) of the Gaia mission contains 1.7 billion sources and
was made public on 24$^{th}$
April 2018 \citep{2016A&A...595A...1G, 2016A&A...595A...2G, 
2017A&C....21...22S, 2010A&A...523A..48J}.
The Gaia DR2 catalogue provides five-dimensional
astrometric data (position, parallax, and proper motion), three-dimensional photometric data and spectroscopic data
\citep{2018A&A...616A..11G}.
The data of Gaia DR2 are based on
observations taken for 22 months \citep{2018A&A...616A...2L}.

Since open clusters orbit near the Galactic plane, they are constantly disturbed 
by tidal forces originating from Galactic disc and the molecular clouds 
present there. Due
to this disturbance and dynamical evolution of the clusters,
the structure of open clusters get changed
\citep{2013MNRAS.432.1672L}. Orbits of open clusters are helpful to understand 
the effect of tides and also formation and evolution processes of clusters.
The proper motion and radial velocity data provided by Gaia DR2 catalogue 
can be very helpful to study the kinematics of open clusters.

In this paper, we present wide-field optical photometry of three
open star clusters namely NGC 6067, NGC 2506 and IC 4651. The main
goal of this article is to investigate the parameters of clusters using 
the member stars selected from proper motion and parallax data. Apart from this, 
we also studied the galactic orbits and dynamical evolution of the
clusters considering their member stars.  

In section \ref{sec:lit}, we present a literature survey for the clusters under study.  
Section \ref{sec:obs} describes observational data, methods of data reduction and calibration. 
The kinematical data are described in section \ref{sec:kin}.
In section \ref{sec:or}, orbits of the clusters are calculated. The basic parameters
of the clusters are discussed in section \ref{sec:ana}.  
Luminosity, mass functions, and mass segregation are described in section \ref{sec:luminosity}.
Results and conclusions 
are presented in section \ref{con}. The basic parameters of the clusters taken from
WEBDA database are listed in Table 
\ref{bpara}.

\begin{table*}
   \centering
   \caption{ General information of the clusters under study, as given in 
    WEBDA database}
   \begin{tabular}{ccccccccc}
   \hline\hline
  Cluster &  $\alpha$ & $\delta$  & $l$ (deg)  & $b$ (deg) & $d_{\odot}$ (pc) & $E(B-V)$ & $log(t)$ & $[Fe/H]$ (dex) \\
  \hline
   NGC 6067 & 16:13:11 &  -54:13:06 & 329.74 & -2.20 & 1417 & 0.38 & 8.08 & +0.13   \\ 
   NGC 2506 & 08:00:01 &  -10:46:12 & 230.56 & 9.93 & 3460 & 0.08 & 9.04 & -0.37   \\
   IC 4651 & 17:24:49 &  -49:56:00  & 340.09 & -7.91 & 888 & 0.11 & 9.06 & +0.1    \\
  \hline
  \end{tabular}
  \label{bpara}
  \end{table*}
% previous studies
%_________________________________________________________________

\section{Previous studies} \label{sec:lit}

{\bf \it NGC 6067}: NGC 6067 is a young open cluster of age $\sim
10^{8}$ years, superimposed on the Norma star cloud. \citet{1962MNRAS.124..445T}
studied this cluster first time and provided radial velocities and 
parallaxes for few stars. \citet{1966ArA.....4...53E} determined cluster distance 
as 2100 pc using photographic data. \citet{1985MNRAS.214...45W} estimated distance modulus 
and reddening as $11.05\pm 0.10$ and 0.35 mag respectively using $BV$ CCD 
photometric data.

{\bf \it NGC 2506}: NGC 2506 is an intermediate age open star cluster
of age $ \sim 10^{9}$ years.
This was first studied by \citet{1981ApJ...243..841M} using 
photoelectric and photographic photometry and derived the age
as $3.4 \times 10^{8}$ years. \citet{1981ApJ...243..827C}
identified four blue stragglers stars in this cluster. 
\citet{1997MNRAS.291..763M} conducted $UBGVRI$ CCD photometry 
and determined the cluster parameters as 
$(m-M)_{0} = 12.6 \pm 0.1$ mag, age $= 1.5 - 2.2$ Gyr and $E(B-V) = 0 - 0.07$ mag.
\citet{2012MNRAS.425.1567L} investigated this cluster using $VI$ CCD data and
calculated $E(B-V) = 0.03 \pm 0.04$ mag, age $= 2.32 \pm 0.16$ Gyr, $\rm {[Fe/H]}$ $
= -0.24 \pm 0.06$ dex and the mass function slope
as $-1.26 \pm 0.07$ in units of log(mass).

{\bf \it IC 4651}: This is an intermediate age open star 
cluster of age $\sim$ $10^{9}$ years. \citet{1971ApJ...166...87E} estimated reddening $E(B-V)=0.15$ mag
using $UBV$ photographic data. \citet{2000A&A...361..929M} have done $uvby$ CCD photometric study 
and estimated distance modulus as $10.03 \pm 0.1$ mag. 
\citet{2000AJ....119.2282A} studied this
cluster using $uvbyH\beta$ CCD photometry and derived the colour excess $E(b-y) = 0.062 \pm 0.003$, 
[Fe/H] $= +0.077 \pm 0.012$ and $(m-M)= 10.15$ mag.
\citet{2007A&A...475..981B} performed spectroscopic study of the cluster
IC 4651 and calculated $E(B-V)$ as $0.12 \pm 0.02$
and age as $1.2 \pm 0.2$ Gyr. 

% data set
%_________________________________________________________________
\section{Observational data and reduction} \label{sec:obs}

\subsection{Optical data} \label{sec:opt}

A large area of the cluster is necessary to study the complete census of the cluster.
The Wide Field Imager (WFI) mounted on ESO/MPI telescope 2.2m at La Silla (Chile) is an ideal instrument to observe 
a large area of the clusters. WFI is a mosaic camera consisting of 
4$\times$2 i.e. 8 CCD chips. Since each CCD has an array of 2048$\times$4096 pixels, WFI ultimately 
produces an image with a 34$\times$33 arcmin$^2$ field of view. Knowing the capabilities 
of WFI, we searched CCD photometric data in ESO archive \footnote{\url{http://archive.eso.org/eso/eso\_archive\_main.html}}
for the three clusters NGC 6067, NGC 2506 and IC 4651. A log of data of the observations is provided 
in Table \ref{log}. In total, 48 images for the cluster NGC 6067 in $B$ and $V$ band, 10 images for the cluster 
NGC 2506 in $B$, $V$ and $I$ band and 8 images for the cluster IC 4651 in $B$, $V$ and $I$ band are found 
suitable for our analysis. Along with science images, many bias and flat images are also taken to pre-reduct
the science images.

%---------------------------------------------------------------------

\subsection{Data reduction and calibration} \label{sec:red}

For the photometric reductions, we followed the procedures 
outlined in \citet{2006A&A...454.1029A}. 
To pre-reduct the science images we adopted the standard 
procedure of bias subtraction and flat-fielding using
$mscred$ package under $IRAF$. An empirical point-spread-function 
(PSF) for each image was constructed 
to find out the position and flux of stars. An array of 15 PSFs 
for each 2048$\times$4096 chip was 
made. This is because of PSF changes with the position in 
the chip \citep{2006A&A...454.1029A}. In this way, 120 PSFs 
were considered for the entire field of view (8192$\times$8192 pixels) 
and saved in a look-up 
table on a very fine grid. To construct the PSF, we selected 
bright and isolated stars with an automatic code.
To determine the positions and magnitude of the stars, an 
iterative procedure is designed to work from brightest 
to the faintest stars. Instrumental magnitudes were determined 
by applying exposure time and extinction 
coefficient (taken from La Silla site) correction for each filter.

\begin{figure}
    \centering
    \includegraphics[width=8cm, height=8cm]{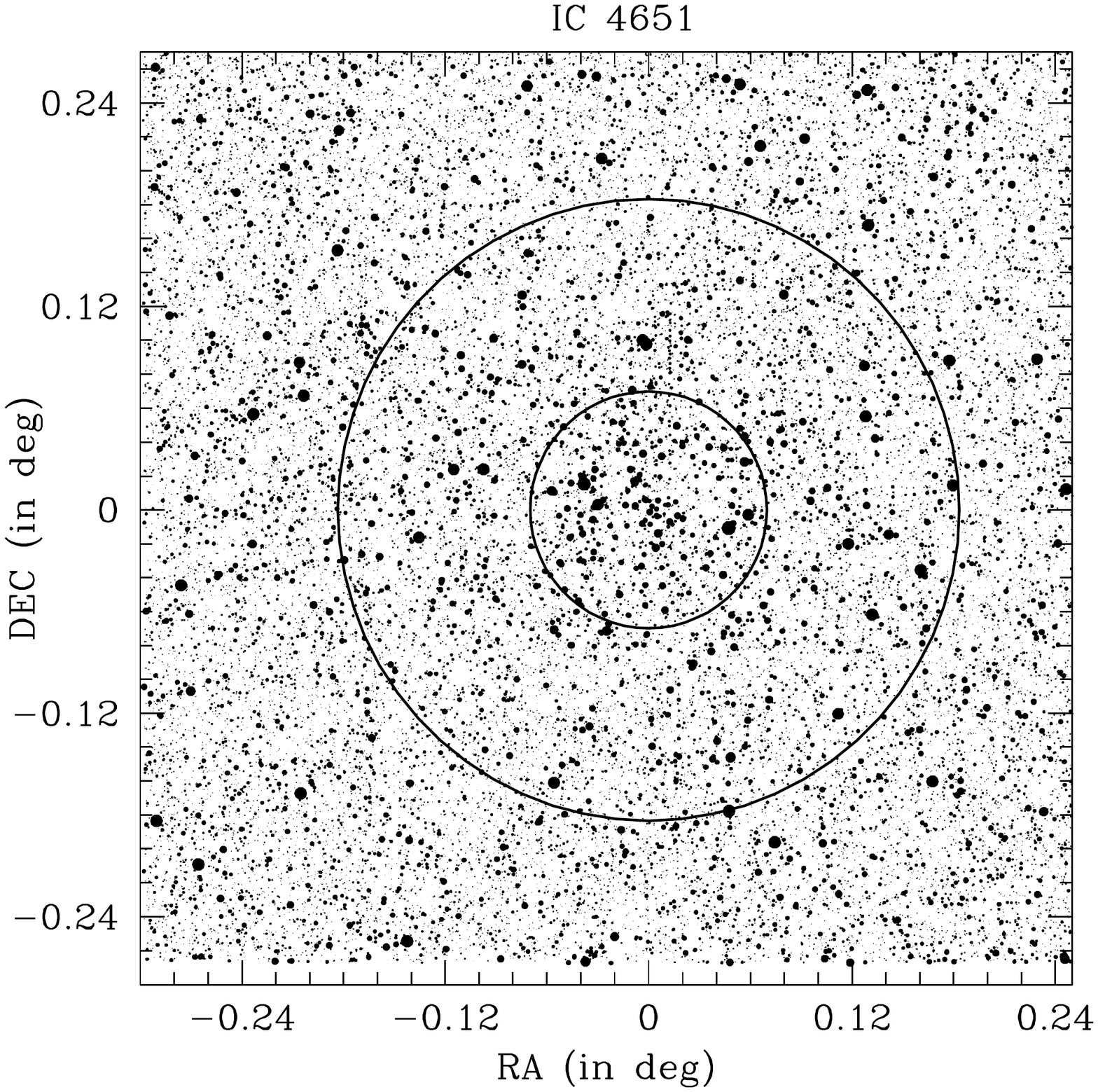}
    \caption{Same as Fig. \ref{id} for the cluster IC 4651. 
     }
    \label{id2}
  \end{figure}

\begin{table}
   \centering
   \caption{Details of data for the clusters.}
   \begin{tabular}{ccc}
   \hline\hline
  Band & Exposure Time & Date   \\
        & (in seconds)    &   \\
  \hline
  &  {\bf NGC 6067}  &    \\
   $B$   & 10 $\times$ 8 & May 1999  \\
       & 60 $\times$ 6 &  "   \\
       & 300 $\times$ 6 &  "    \\
       & 360 $\times$ 1 &  "    \\
   $V$   & 10 $\times$ 6  &  "     \\
       & 60 $\times$ 6  &  "     \\
       & 300 $\times$ 8 &   "    \\
       & 360 $\times$ 7 &   "    \\
   & {\bf NGC 2506}   &   \\
   $B$   & 30 $\times$ 1  & Nov 2000    \\
       & 240 $\times$ 2 &  "   \\
   $V$   & 30 $\times$ 1  &  "   \\
       & 60 $\times$ 1  &  Dec 2012   \\
       & 240 $\times$ 1 &  Nov 2000 \\
   $I$   & 30 $\times$ 2  &  "   \\
       & 240 $\times$ 2 &  "   \\
   & {\bf IC 4651}  &  \\
  $B$    & 20 $\times$ 1  &   June 2000   \\
       & 240 $\times$ 2 &     "      \\
  $V$    & 240 $\times$ 2 &     "       \\
  $I$    & 240 $\times$ 3 &     "       \\
  \hline
  \end{tabular}
  \label{log}
  \end{table}

  \begin{table*}
   \centering
   \caption{ The colour coefficients ($C_{X}$) and zero-points
             ($Z_{X}$) for respective filters
           used for the calibration equations. $X$ represents different filter systems.
           }
   \begin{tabular}{ccccccccc}
   \hline\hline
  Cluster & $C_{b}$ & $C_{v}$ & $C_{i}$ & $Z_{b}$  & $Z_{v}$ & $Z_{i}$ \\
  \hline
   NGC 6067  &  $0.49\pm 0.031$ &  $ -0.12 \pm 0.017 $ & - & $24.92 \pm 0.007$ & $24.02 \pm 0.004$ & -   \\
   NGC 2506  & $0.37\pm 0.030  $  & $ -0.22\pm 0.017 $ & $0.03 \pm 0.019  $ & $24.69 \pm 0.003 $ & $ 23.95\pm 0.002$ & $20.86 \pm 0.045 $     \\
   IC 4651   & $0.16 \pm 0.045 $ & $-0.11 \pm 0.015$ & $0.11 \pm 0.028$  & $24.47 \pm 0.008$ & $23.83 \pm 0.005 $ & $23.00 \pm 0.009$   \\
  \hline
  \end{tabular}
  \label{zp}
  \end{table*}

  \begin{figure}
    \centering
    \includegraphics[width=9cm]{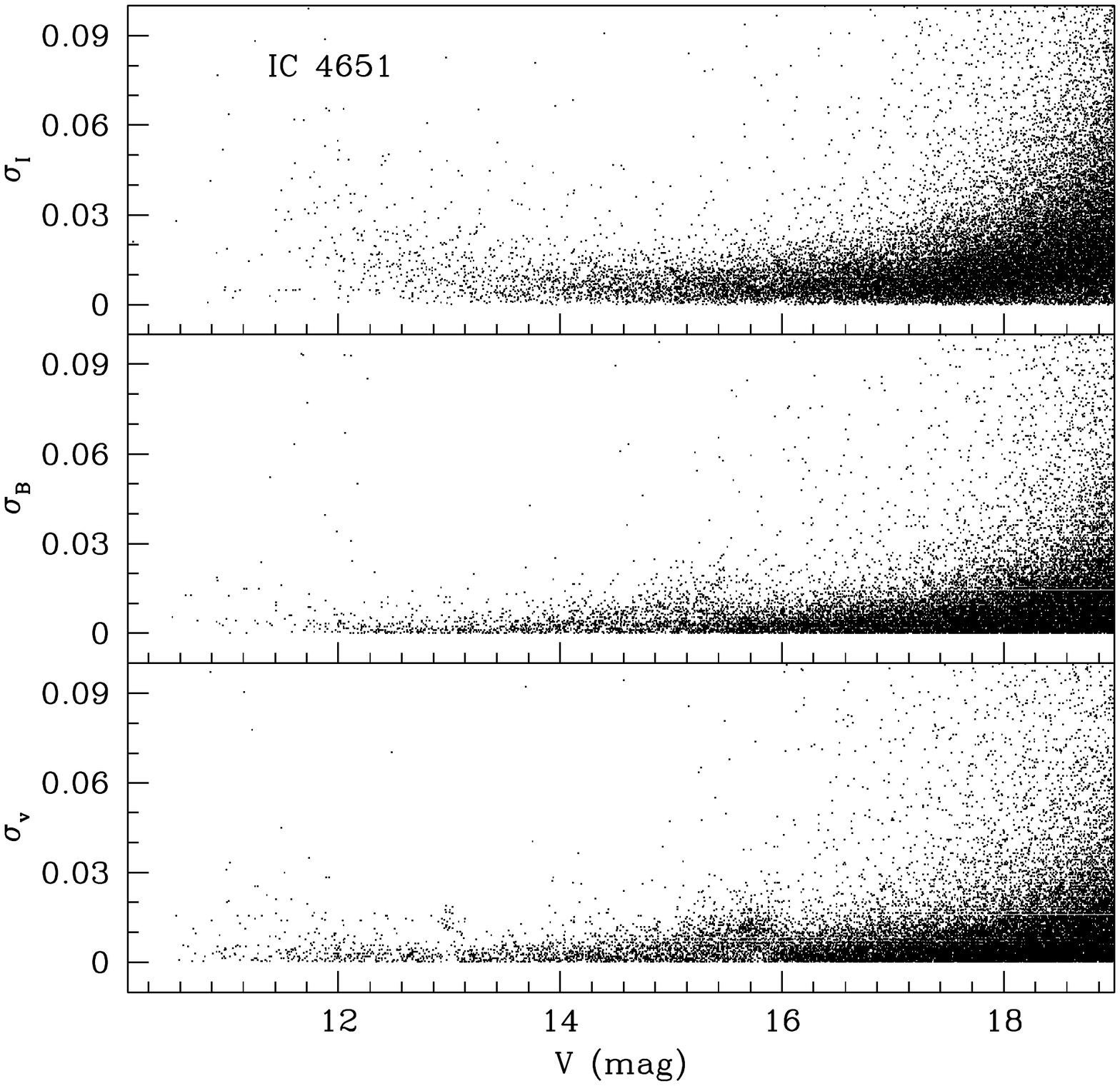}
    \caption{The photometric errors $\sigma_B$, $\sigma_V$ and
     $\sigma_I$ are plotted against the calibrated $V$ magnitude for the cluster IC 4651.}
    \label{erb}
  \end{figure}

\begin{figure}
    \centering
    \includegraphics[width=9cm]{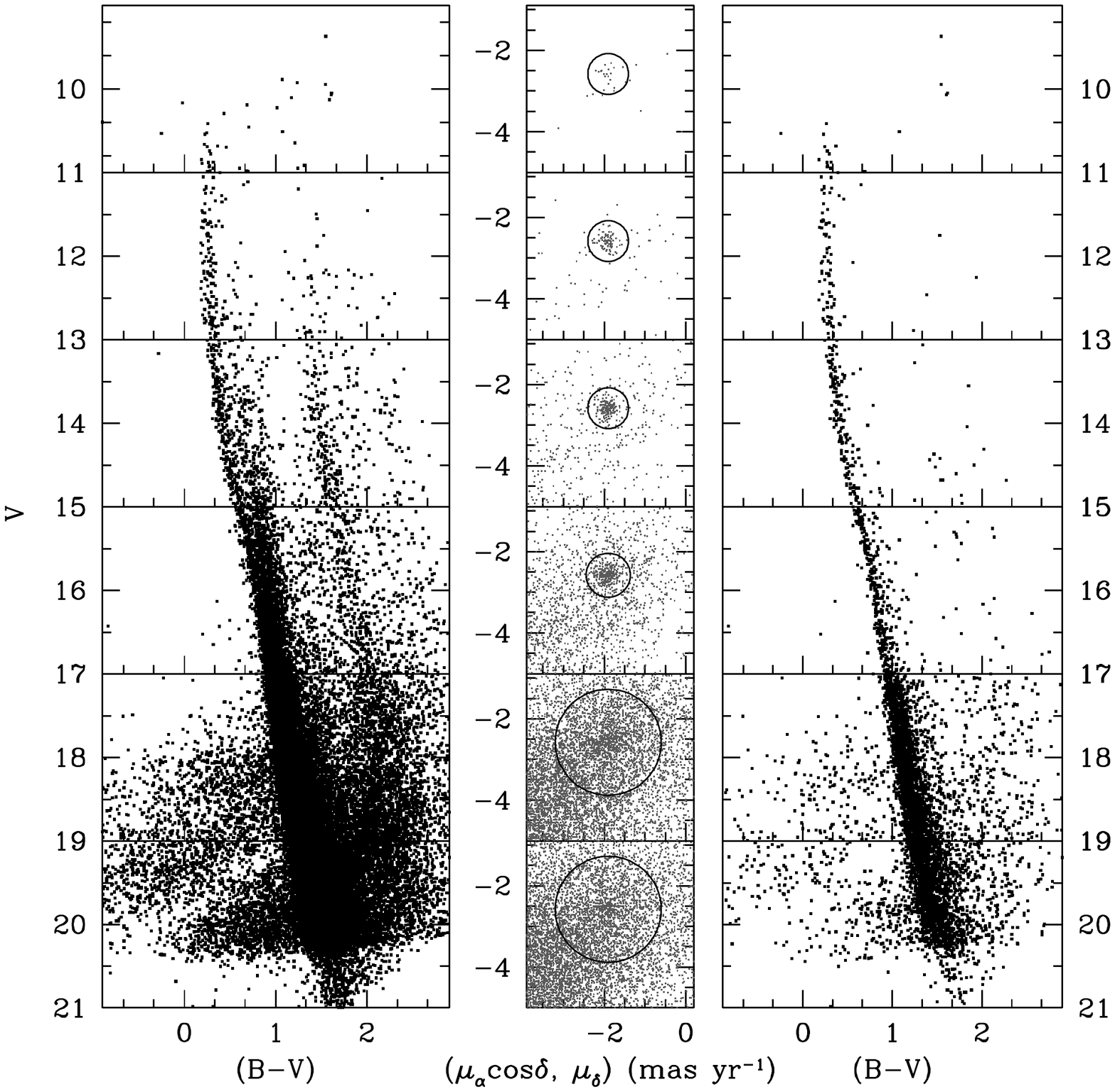}
    \caption{ Proper motion vector-point diagram (middle panels) in 
     different magnitude bins and
     calibrated $(B - V)$,
     $V$ CMD (left and right panels) are shown for the cluster NGC 6067. Left panel shows
     the entire cluster stars and the right panel shows stars in vector
     point diagram within
     the chosen circle radius as stated in the text.
     }
    \label{vpd6}
  \end{figure}
%------------------------------------------------------------------------

%\subsection{Calibration} \label{sec:cal}

   To transform the $BVI$ instrumental magnitudes into Johnson and
Kron-Cousin standard magnitudes, we used the photometric data available in 
the WEBDA database. For NGC 6067, photoelectric $BV$ data by 
\citet{1962MNRAS.124..445T} was used. CCD $BV$ photometric data from 
\citet{1997MNRAS.291..763M} and $VI$ data from \citet{2001AcA....51...49K} was considered for NGC 2506 calibration. For IC 4651, 
CCD $BV$ data from \citet{1991A&AS...87..119K} and $VI$ data from \citet{1998yCat..21160263P} 
were used.\\
 The transformation equations used for calibration are as follows \\

$B_{std} = B_{ins} + C_{b} \times (B_{ins} - V_{ins})$ + $Z_b$ \\

$V_{std} = V_{ins} + C_{v} \times (B_{ins} - V_{ins})$ + $Z_v$  \\

$I_{std} = I_{ins} + C_{i} \times (V_{ins} - I_{ins})$ + $Z_i$ \\

where $B_{std}$, $V_{std}$, $I_{std}$ and $B_{ins}$, $V_{ins}$, $I_{ins}$ 
are the standards and instrumental magnitudes
while $C_{b}$, $C_{v}$, and $C_{i}$ are the colour coefficients 
for $B$, $V$, and $I$ filters respectively. $Z_{b}$, $Z_{v}$ 
and $Z_{i}$ are the zero-points for respective filters. 
The colour coefficients and zero-points obtained during transformation 
for each cluster are listed in Table \ref{zp}. We converted $X$ and $Y$ 
position of stars into the right ascension (RA) and declination (DEC) of $J2000$ 
using $CCMAP$ and $CCTRAN$ tasks in IRAF. The standard error in transformation 
is $\sim$ 75 mas in each coordinate. The identification map for the clusters 
under study are shown in Figures \ref{id} and \ref{id2}. Inner and outer 
circles show the core and cluster radius derived in the present study.

The rms of the residuals around the mean magnitude as a function of 
$V$ magnitude for each filter is shown in Fig. \ref{erb} for IC 4651. The photometric standard 
deviations are computed using multiple observations and then
reduced to a common reference frame. Stars with $V \lesssim 12$ mag have higher 
dispersion because of saturation. The average internal errors in our photometry
in each filter are listed in Table \ref{ert}. On average, photometric errors are 
lower than $\sim$ 0.01 mag for stars with $V \lesssim 20$ mag.

\begin{table}
   \tiny
   \centering
   \caption{ Internal photometric errors as a function of $V$ mag.
   Here $\sigma$ is the mean photometric error in particular mag bin.}
   \begin{tabular}{ccccccccc}
   \hline\hline
  V  & \multicolumn{2}{c|}   {NGC 6067} & \multicolumn{3}{c} {NGC 2506}  & \multicolumn{3}{c}  {IC 4651}  \\
(mag)      & $\sigma_{B}$  & $\sigma_{V}$ & $\sigma_{B}$  & $\sigma_{V}$ & $\sigma_{I}$  & $\sigma_{B}$  & $\sigma_{V}$ & $\sigma_{I}$   \\
  \hline
   12 - 14 & 0.024  &  0.032  &  0.015  &  0.008   &  0.031  & 0.006  &  0.017 &  0.014   \\
   14 - 16 & 0.014  &  0.018  &  0.010  &  0.005   &  0.019 & 0.006  &  0.017 &  0.014  \\
   16 - 18 & 0.010  &  0.012  &  0.006  &  0.006   &  0.010 & 0.006  &  0.014 &  0.009    \\
   18 - 20 & 0.011  &  0.012  &  0.006  &  0.008   &  0.010 & 0.008  &  0.015 &  0.012   \\

  \hline
  \end{tabular}
  \label{ert}
  \end{table}

%proper motion data
%---------------------------------------------------------------------
\section{Kinematical data and mean proper motion of clusters} \label{sec:kin}

Proper motion data are very valuable to distinguish cluster members from field 
stars and also to study the kinematics of open star clusters. The availability 
of such kind of data in Gaia DR2 catalogue has become very useful. Therefore,
we retrieved kinematical data from the Gaia DR2 database \citep{2018A&A...616A...1G}
for the clusters under study. We extracted all the Gaia DR2 sources within
a radius of 15$^\prime$ from the cluster centre of each cluster to get
as complete as possible sample of cluster stars. 
The limiting magnitude for Gaia DR2 data is 
$G \simeq 21.0$ mag. Errors in trigonometric parallax are $\sim 0.04$ mas for
sources having $G \sim 14 $ mag, $\sim 0.1$ mas for $G \sim 17 $ mag and
$\sim 0.7$ mas for $G \sim 20 $ mag \citep{2018A&A...616A...2L, 2018A&A...616A...9L}.
Proper motion uncertainty is
0.2 mas $yr^{-1}$ for stars brighter than $G \sim 17 $ mag
\citep{2018A&A...616A...2L}.
The average errors of proper motions 
in different magnitude bins for the three clusters are 
listed in Table \ref{pert}. The average errors are $\sim$ 0.07 mas $yr^{-1}$ in $\mu_{\alpha}cos\delta$ 
and $\sim$ 0.05 mas $yr^{-1}$ in $\mu_{\delta}$.

\begin{table}
   \tiny
   \centering
   \caption{The errors in the Gaia DR2 proper motion data 
      as a function of G magnitude for the clusters
   NGC 6067, NGC 2506 and IC 4651}
   \begin{tabular}{c|cc|cc|cccc}
   \hline\hline
   G   & \multicolumn{2}{c|} {NGC 6067}  & \multicolumn{2}{c} {NGC 2506}  &\multicolumn{2}{c}  {IC 4651}   \\
  (mag) &  $\sigma_{\mu_{\alpha} cos \delta}$  & $\sigma_{\delta}$ & $\sigma_{\mu_{\alpha} cos \delta}$  & $\sigma_{\delta}$ & $\sigma_{\mu_{\alpha} cos \delta}$  & $\sigma_{\delta}$    \\
  \hline
   10 - 12 &  0.067  &  0.057  &  0.066  &  0.046  &  0.091  &  0.085     \\
   12 - 14 &  0.072  &  0.061  &  0.065  &  0.047  &  0.082  &  0.077     \\
   14 - 16 &  0.054  &  0.046  &  0.049  &  0.035  &  0.062  &  0.059     \\
   16 - 18 &  0.064  &  0.054  &  0.056  &  0.040  &  0.072  &  0.069     \\
   18 - 20 &  0.069  &  0.066  &  0.053  &  0.051  &  0.073  &  0.071     \\
  \hline
  \end{tabular}
  \label{pert}
  \end{table}

% proper motion
%-----------------------------------------------------------------------

\subsection{Selection of cluster members and mean proper motion} \label{sec:vpds}

   To derive the reliable parameters of a cluster, it is very important
to use genuine cluster members. The stars of a cluster 
share the same kinematical properties in the sky and have the same heliocentric distance.
Because of this, proper motion and parallax are among the most reliable 
parameter to differentiate cluster
members from the field stars. 
We used Gaia DR2
proper motion data and parallax data 
to select members and determine their mean proper motion. For this 
purpose, we matched our photometric catalogue with Gaia data and made a catalogue 
of common stars.

\begin{figure}
    \centering
    \includegraphics[width=9cm]{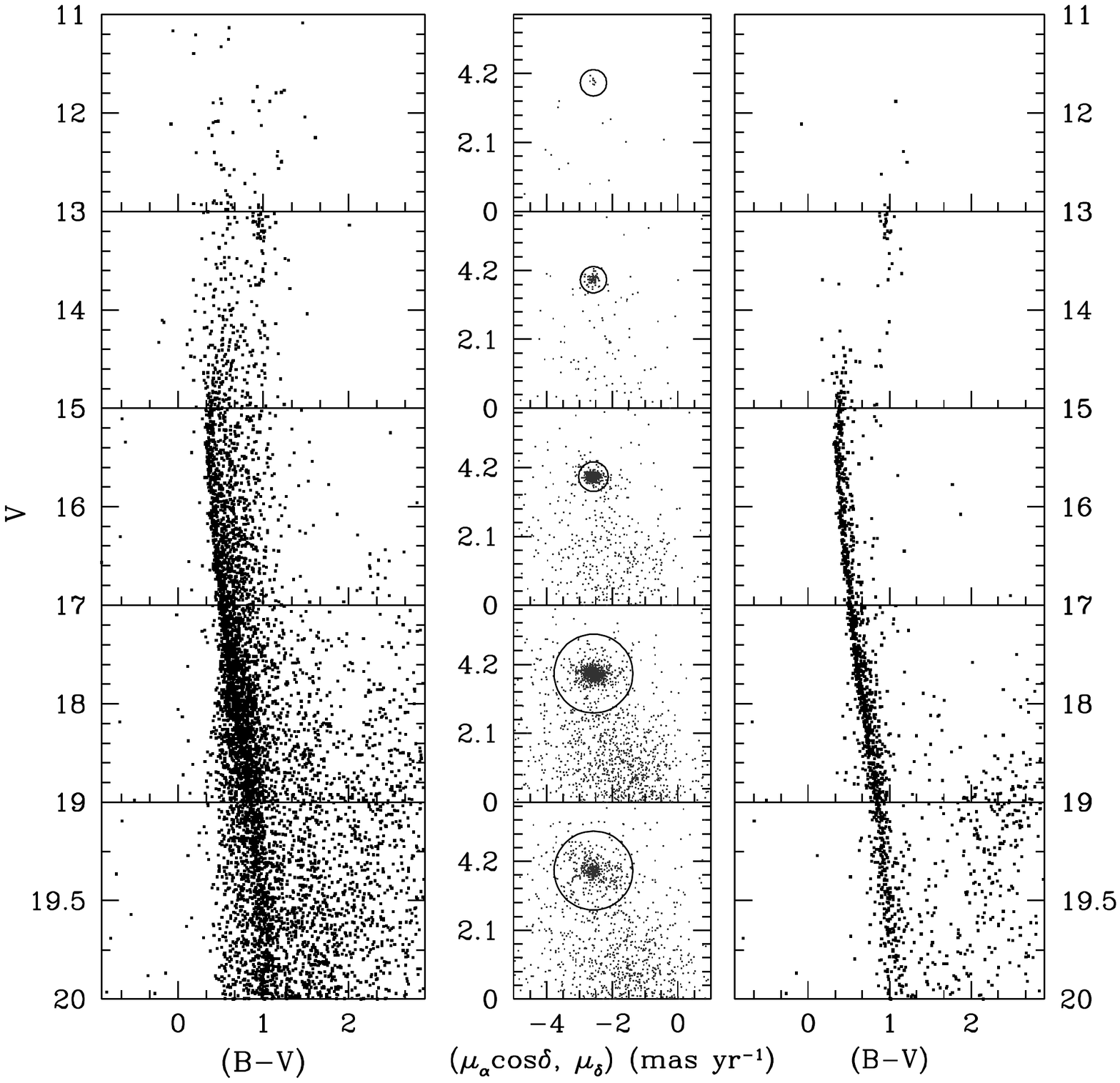}
    \caption{Same as Fig. \ref{vpd6} for NGC 2506.
       }
    \label{vpd2}
  \end{figure}

\begin{figure}
    \centering
    \includegraphics[width=9cm]{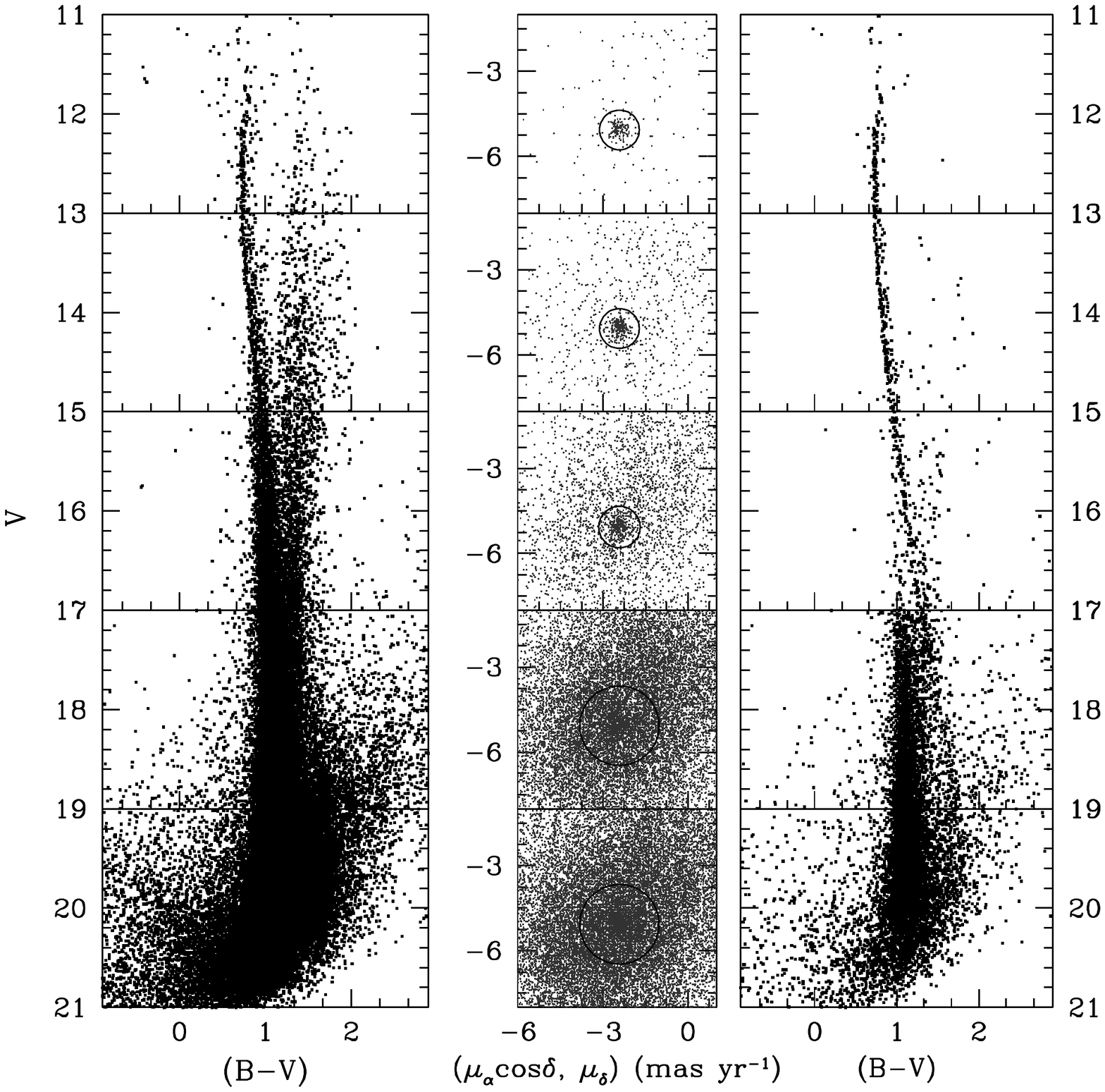}
    \caption{Same as Fig. \ref{vpd6} for IC 4651.
       }
    \label{vpd4}
  \end{figure}

To see the distribution of cluster and field stars, we 
plotted vector point diagrams (VPD) in $\mu_{\alpha}cos\delta$ and $\mu_{\delta}$.
The vector point diagrams for the clusters NGC 6067,
NGC 2506 and IC 4651 are shown in the middle panels of Fig. \ref{vpd6},
\ref{vpd2} and \ref{vpd4} respectively along with $V$ versus $(B-V)$
colour-magnitude diagrams (CMD) in the left and right panels.
The left panels display a complete sample of stars while
right panels show the stars that based on their proper motion,
have a high probability to be a cluster member.
Vector point diagrams of the clusters show two distinct populations for
each cluster. Clusters population are tightly distributed while field populations 
are scattered.

By visual inspection, we define the centre and radius of the cluster
members' proper motion distribution in VPD as shown in 
 \ref{vpd6}, \ref{vpd2} and \ref{vpd4}. This selection was performed
in a way to minimize the field contamination and to save as much as
possible faint cluster members.
The chosen circle radii are 0.5, 0.4 
and 0.7 mas $yr^{-1}$ for NGC 6067, NGC 2506 and IC 4651 respectively.
As expected, the fainter stars have larger measurement errors, therefore 
we selected stars with different circle radius in different magnitude 
bins as shown in the Figs. \ref{vpd6}, \ref{vpd2} and \ref{vpd4}.
The new circle radius is chosen as $r= \sqrt{r_{0}^{2}+\sigma^{2}}$,
where $r_{0}$ is the eye estimated radius and $\sigma$ is the proper motion
error in the corresponding magnitude bin taken from \citet{2018A&A...616A...1G}.

In the right panels, 
the colour-magnitude diagrams of the stars
located within the circles are plotted. These colour-magnitude diagrams
show a clear sequence for all the 
three open star clusters under study. 
It is the first time when the main sequence of IC 4651 is resolved up to
$\sim 19^{th}$ mag. However, some field stars are still visible in the lower 
part of the colour-magnitude diagram of IC 4651. This is because the distribution of these field stars is the same 
as cluster stars in the vector point diagram.

After selecting stars with vector point diagrams,
we again selected members using mean parallax of the cluster.
Parallax angles of stars in the clusters
are available in Gaia DR2 catalogue.
We calculated the 
arithmetic mean of the parallaxes for the stars inside the circle of 
vector point diagram having $V$ mag brighter than 17.0 mag. 
A star is considered as a most probable cluster member if it lies
inside the circle in vector point diagram and 
has a parallax angle within $3\sigma$
from the mean cluster parallax.
In this way, we obtained a total
2894, 1773 and 1947 number of most probable cluster members
for the clusters NGC 6067, NGC 2506 and IC 4651 respectively.
The colour-magnitude diagrams for most probable cluster members are shown in Fig. 
\ref{iso6067},\ref{iso2506} and \ref{iso4651}. A clean main sequence 
is visible in all the clusters under study.
However, few field stars are still present in the colour-magnitude diagrams.
Only the most probable cluster members obtained here are used for 
further investigations. 

\begin{figure}
    \centering
    \includegraphics[width=9cm]{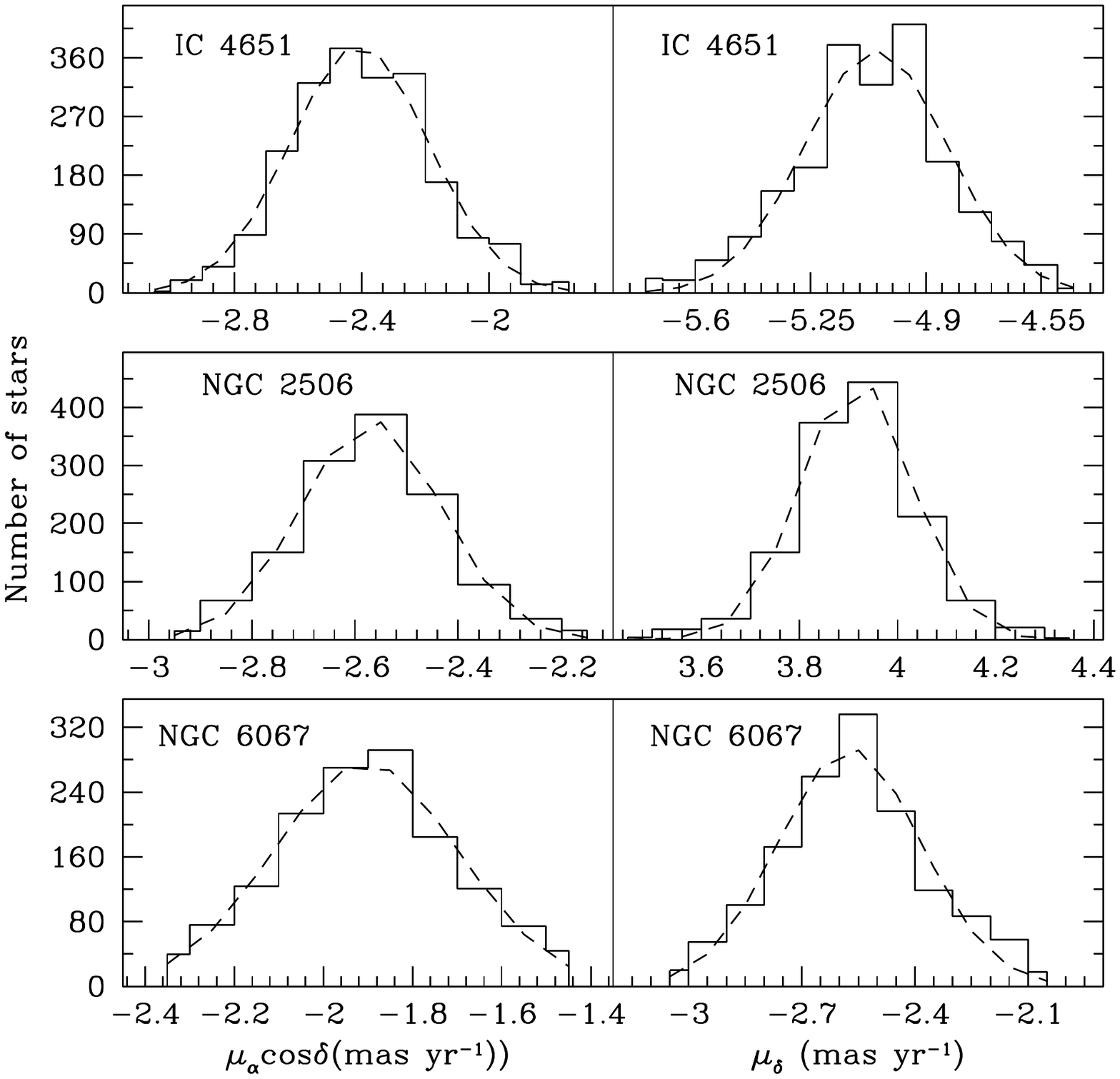}
    \caption{ Histograms in $\mu_{\alpha}cos\delta$ and $\mu_{\delta}$
    for the clusters NGC 6067, NGC 2506 and IC 4651.
     The dashed curves represent Gaussian fitting.
       }
    \label{pra}
  \end{figure}

To calculate mean proper motions in $\mu_{\alpha}cos\delta$ and $\mu_{\delta}$, we plotted
the histograms which are shown in Fig. \ref{pra} for the clusters under study.
The histograms are made by using stars brighter than $V=$ 17 mag. 
A Gaussian function is fitted on the histograms to estimate
the mean and standard deviation in $\mu_{\alpha}cos\delta$ and $\mu_{\delta}$.
In this way, the mean proper motions of the clusters are estimated as
$ -1.90 \pm 0.01$ and $-2.57 \pm 0.01$ mas $yr^{-1}$ for NGC 6067;
$-2.57 \pm 0.01 $ and $3.92 \pm 0.01$ mas $yr^{-1}$ for NGC 2506 and 
$-2.41 \pm 0.01$, 
and $-5.05 \pm 0.02$ mas $yr^{-1}$ for IC 4651.

% Distance and mean proper motion of clusters
%--------------------------------------------------------------------

\subsection{The distance of the clusters using parallax angle} \label{sec:parallax}

\begin{figure}
    \centering
    \includegraphics[width=9cm]{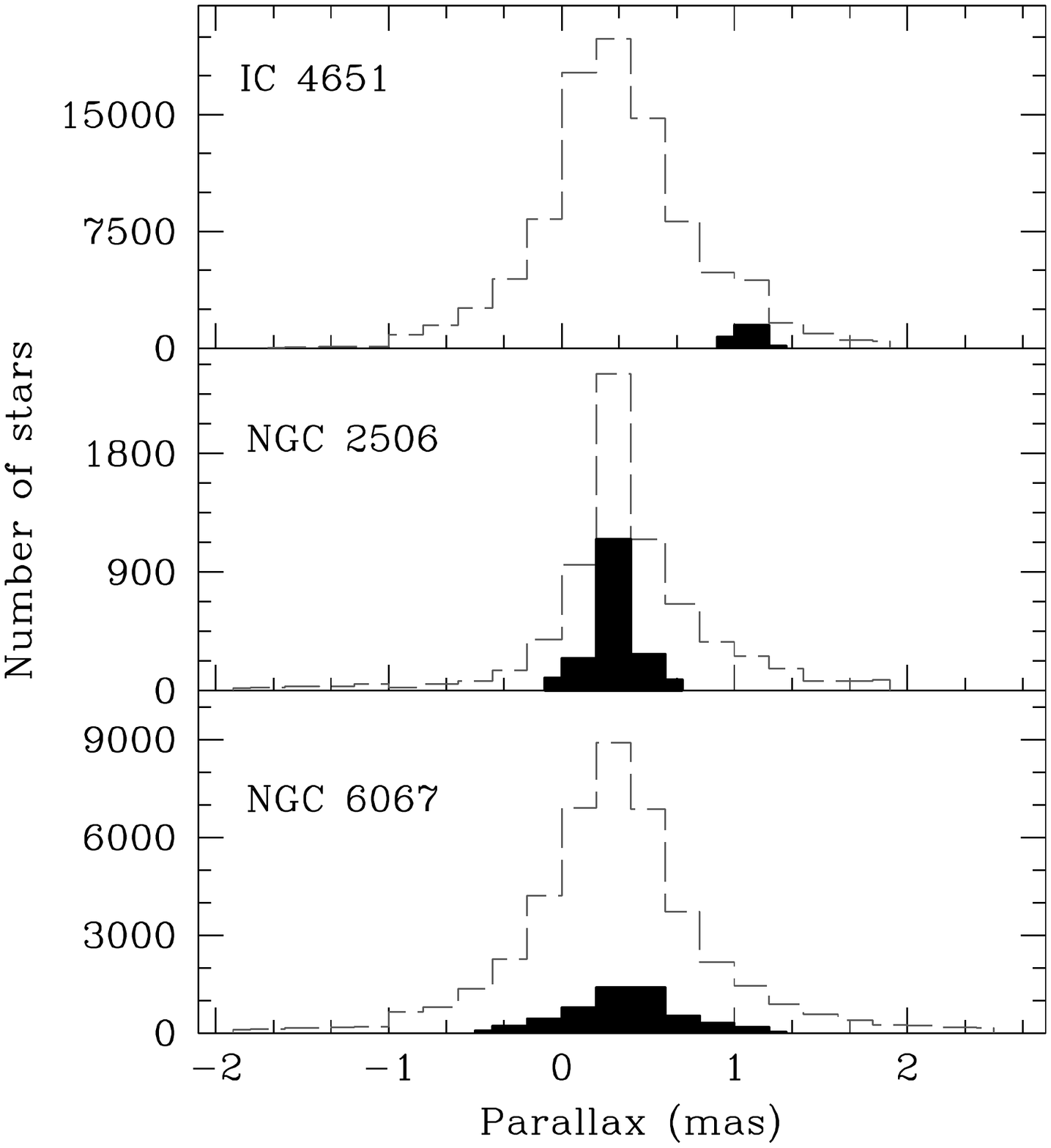}
    \caption{ Histograms of parallax
    for the clusters NGC 6067, NGC 2506 and IC 4651. The dashed
    grey lines represent all stars in the cluster field whereas
    the filled boxes represent the most probable
    cluster members.
       }
    \label{plx}
  \end{figure}

According to \citet{2018A&A...616A...9L} the correct approach to determine the average distance 
of a cluster is to average the parallaxes of cluster stars first and then obtain
distance from the average parallax value.
The parallax data in Gaia DR2 have their
associated errors, so we determined the
weighted mean of the parallaxes for most probable cluster members.
After correcting for Gaia DR2 parallax zero-point (-0.046 mas)
offset as suggested by \citet{2018ApJ...861..126R},
the mean parallax angles are found to be
$0.33 \pm 0.21$, $0.25 \pm 0.10 $ and $1.00 \pm 0.07$ mas
for the clusters NGC 6067, NGC 2506 and IC 4651 respectively.
In Fig. \ref{plx}, we present parallax distributions
of total stars in the cluster region and most probable cluster members,
which are represented by the
dashed gray line and filled boxes respectively.
Histograms show that member stars for the clusters NGC 6067 and IC 4651
are more in comparison to NGC 2506.

\citet{2015PASP..127..994B} have shown that the distance estimated by 
inverting the parallax is not reliable and give incorrect error
estimates. The correct approach to determine distance 
and their uncertainties from the 
parallax are through a probabilistic analysis as described in \citet{2015PASP..127..994B}. 
By adopting the method described in \citet{2018AJ....156...58B}, the distances are determined as
$ 3.01 \pm 0.87 $, $3.88 \pm 0.42 $ and $1.00 \pm 0.08 $ kpc
for the cluster NGC 6067, NGC 2506 and IC 4651 respectively.
Mean parallax angles and corresponding heliocentric
distances of the clusters are also listed in Table \ref{vinp}.
An offset of 0.61 mas between the mean parallax of most probable members and all
the stars in the field of IC 4651 is observed. We estimated the distance of non-probable 
cluster members of IC 4651 using the parallax distribution of field stars and found to be $5.48 \pm 1.54$ kpc. 
A comparison of this distance with the cluster's distance implies that non-members are located 
behind the cluster IC 4651.

\begin{figure}
    \centering
    \includegraphics[width=4cm,height=4cm]{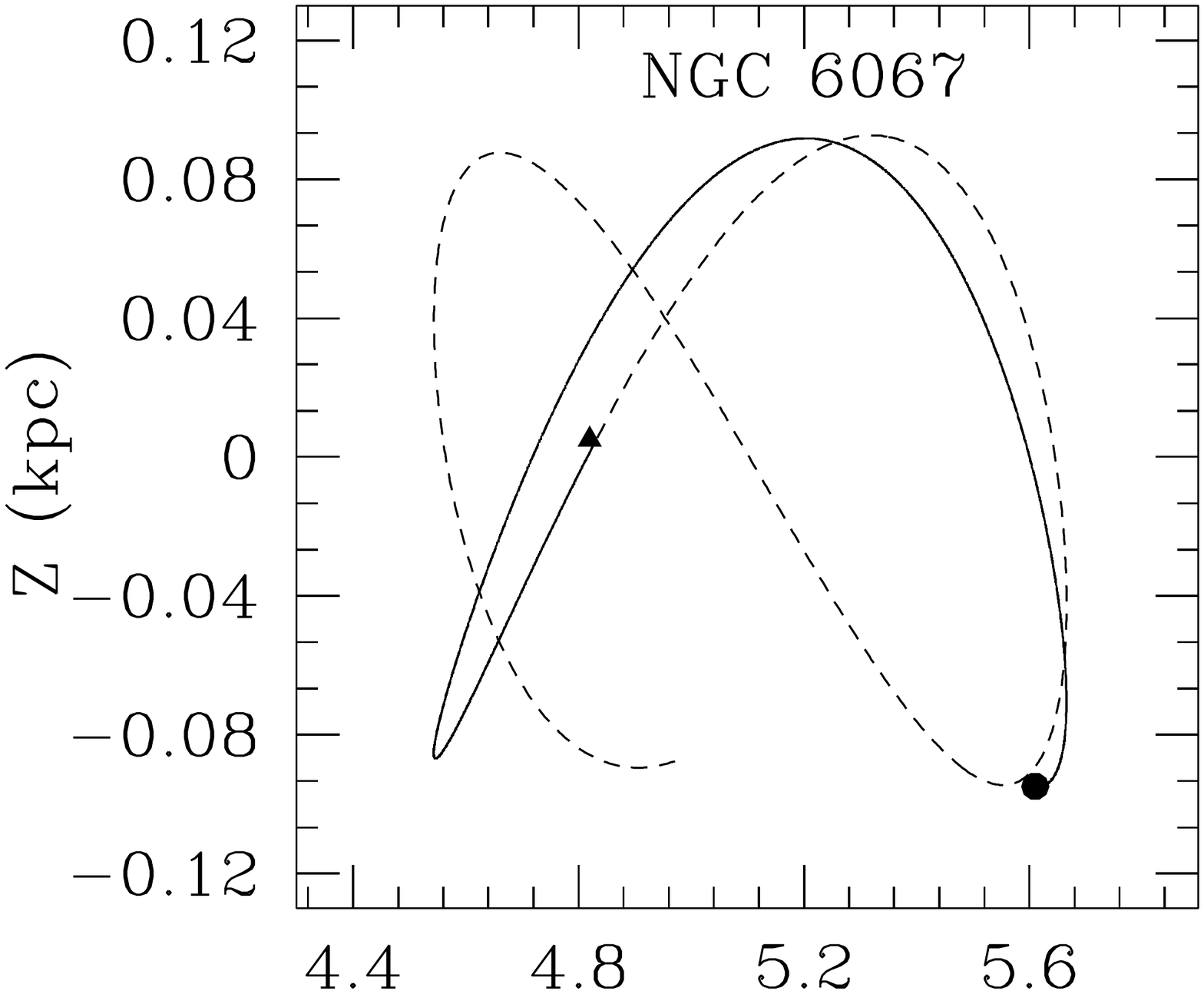}
    \includegraphics[width=4cm,height=4cm]{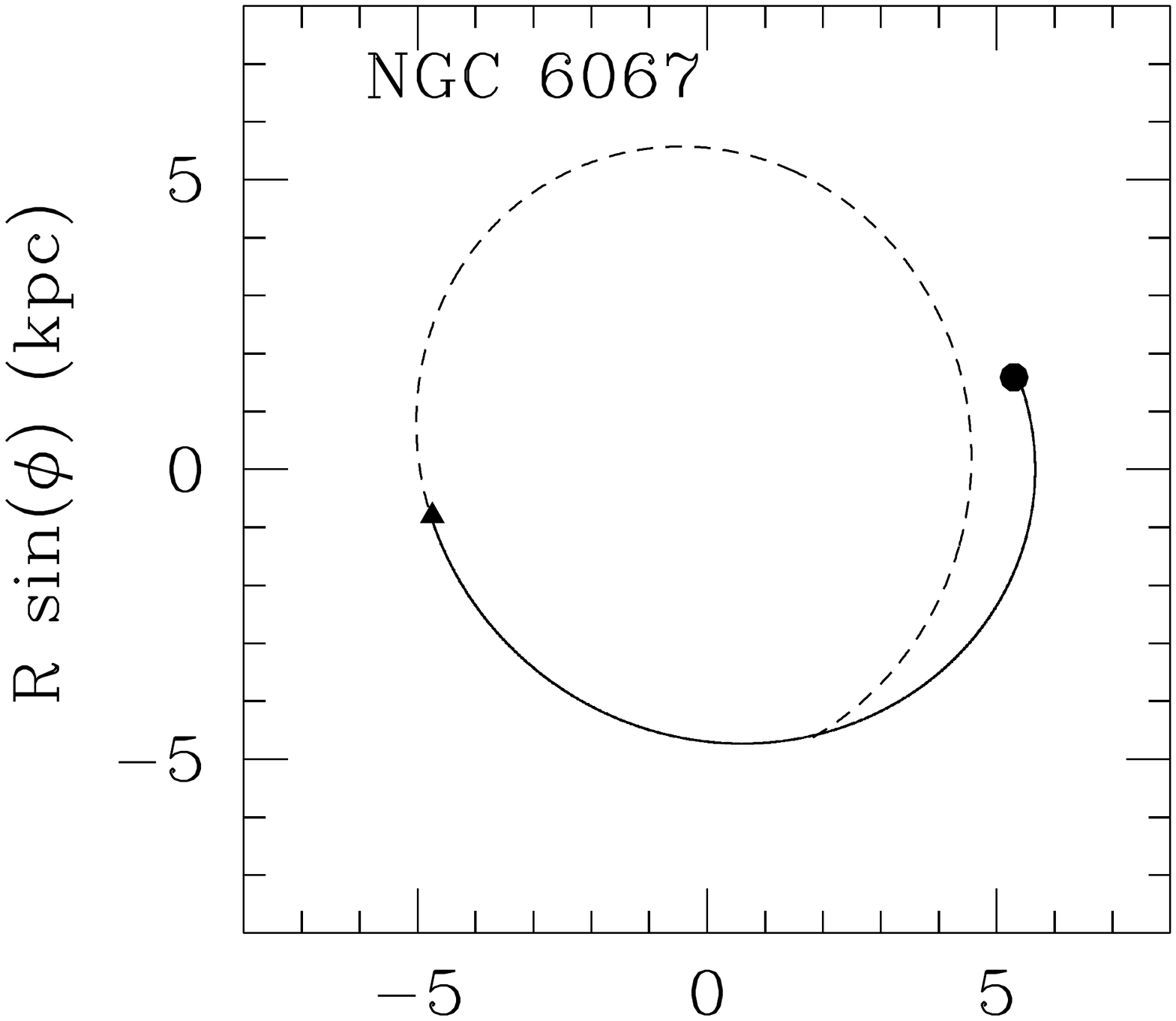}
    \includegraphics[width=4cm,height=4cm]{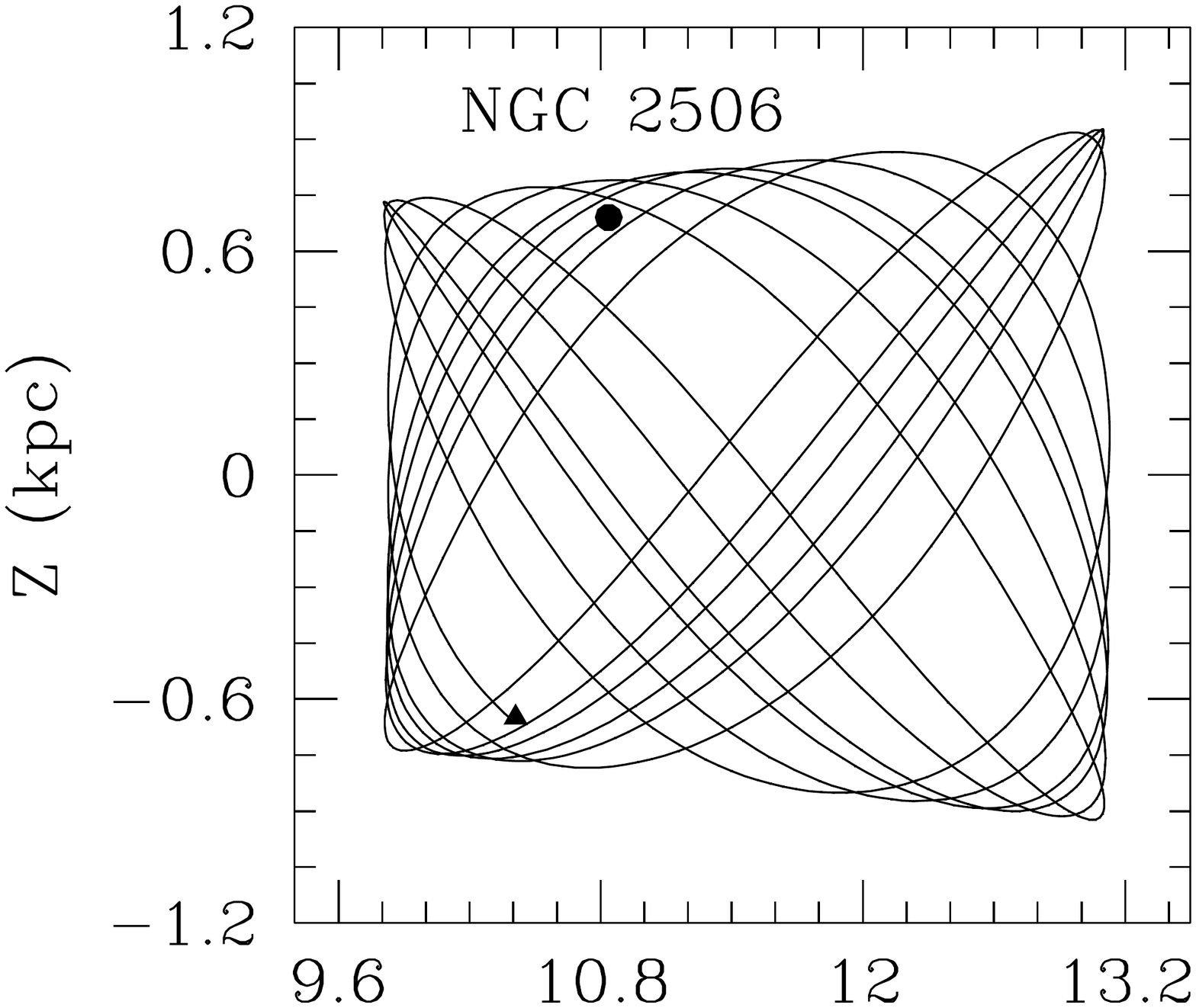}
    \includegraphics[width=4cm,height=4cm]{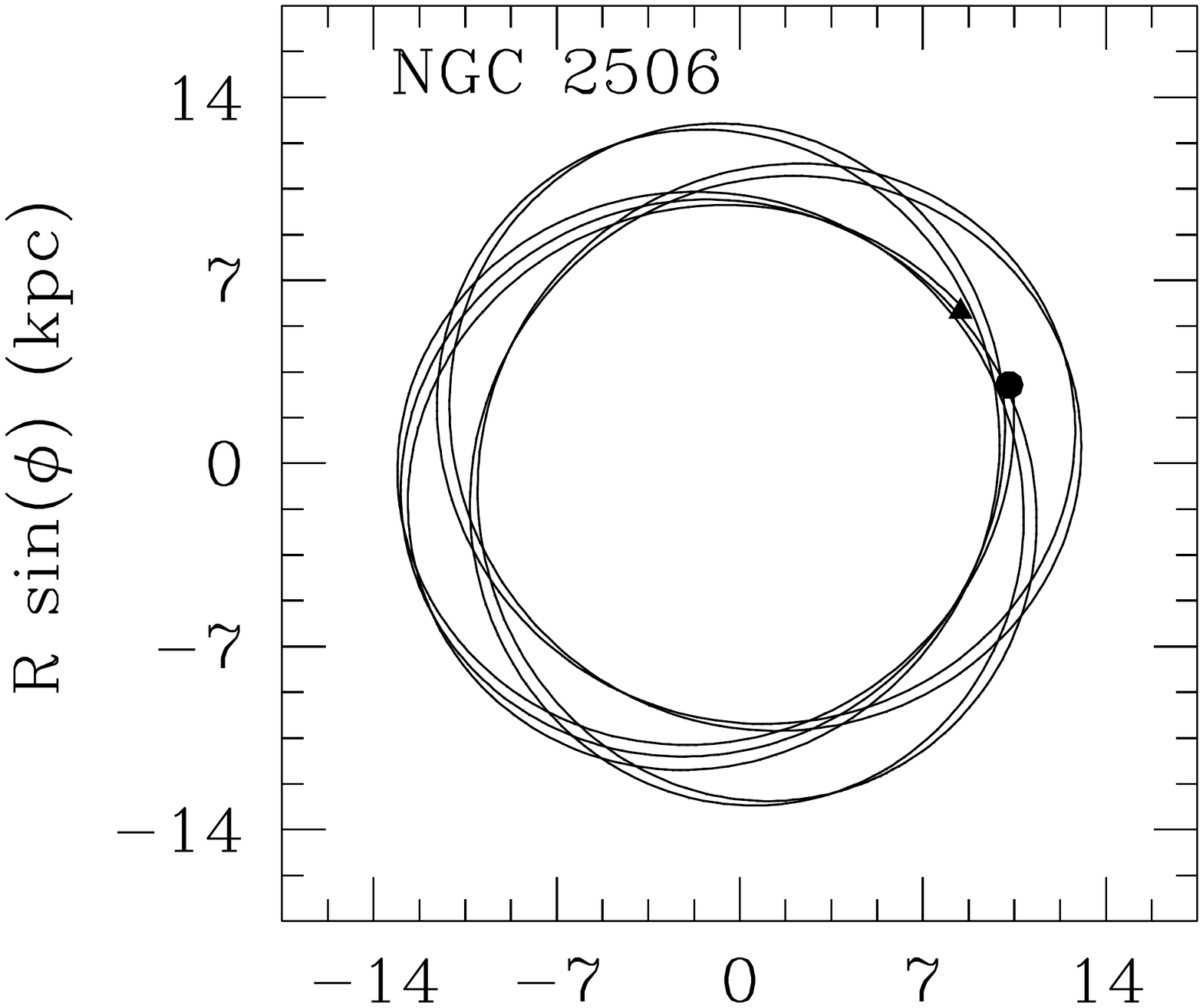}
    \includegraphics[width=4cm,height=4cm]{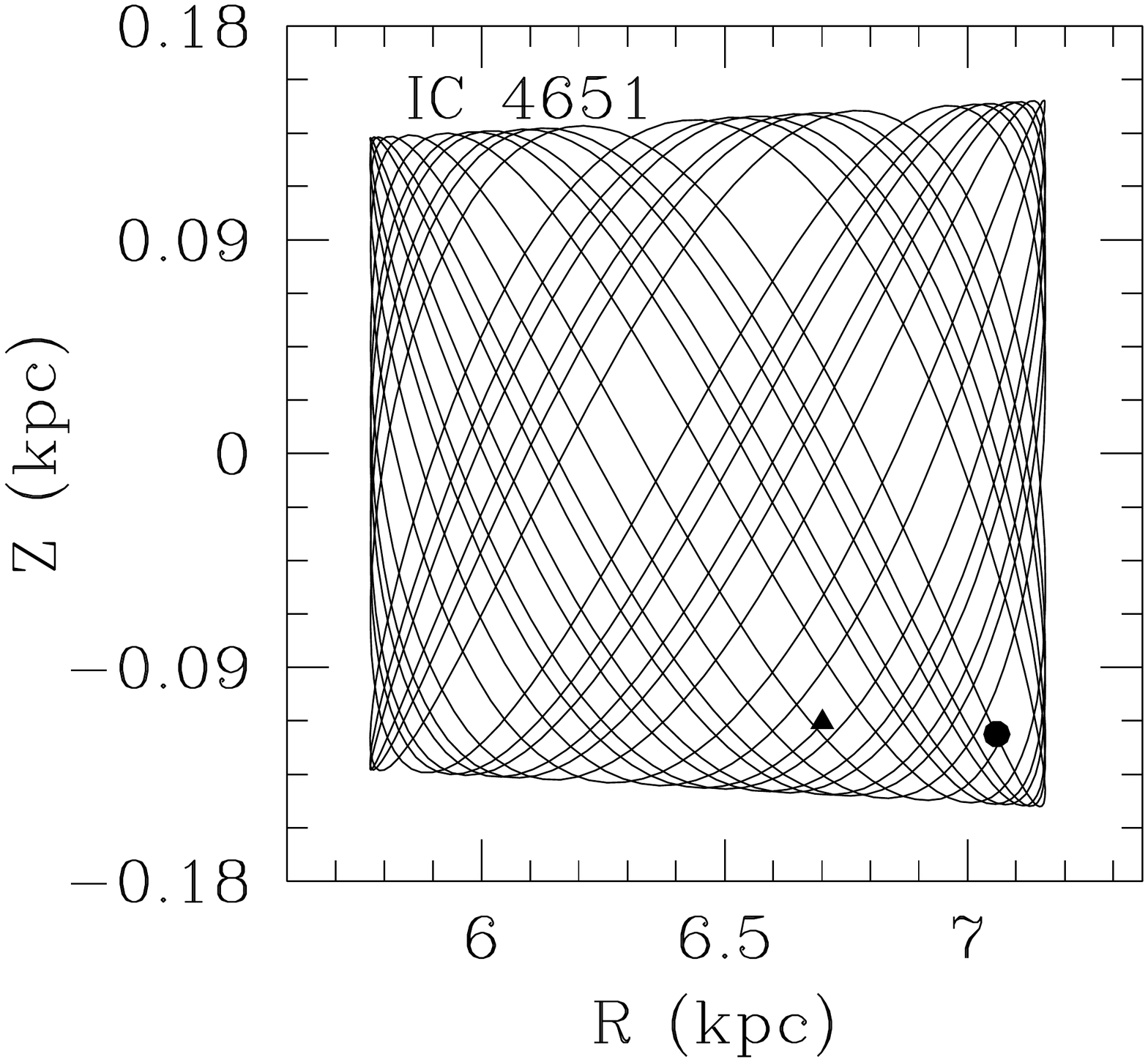}
    \includegraphics[width=4cm,height=4cm]{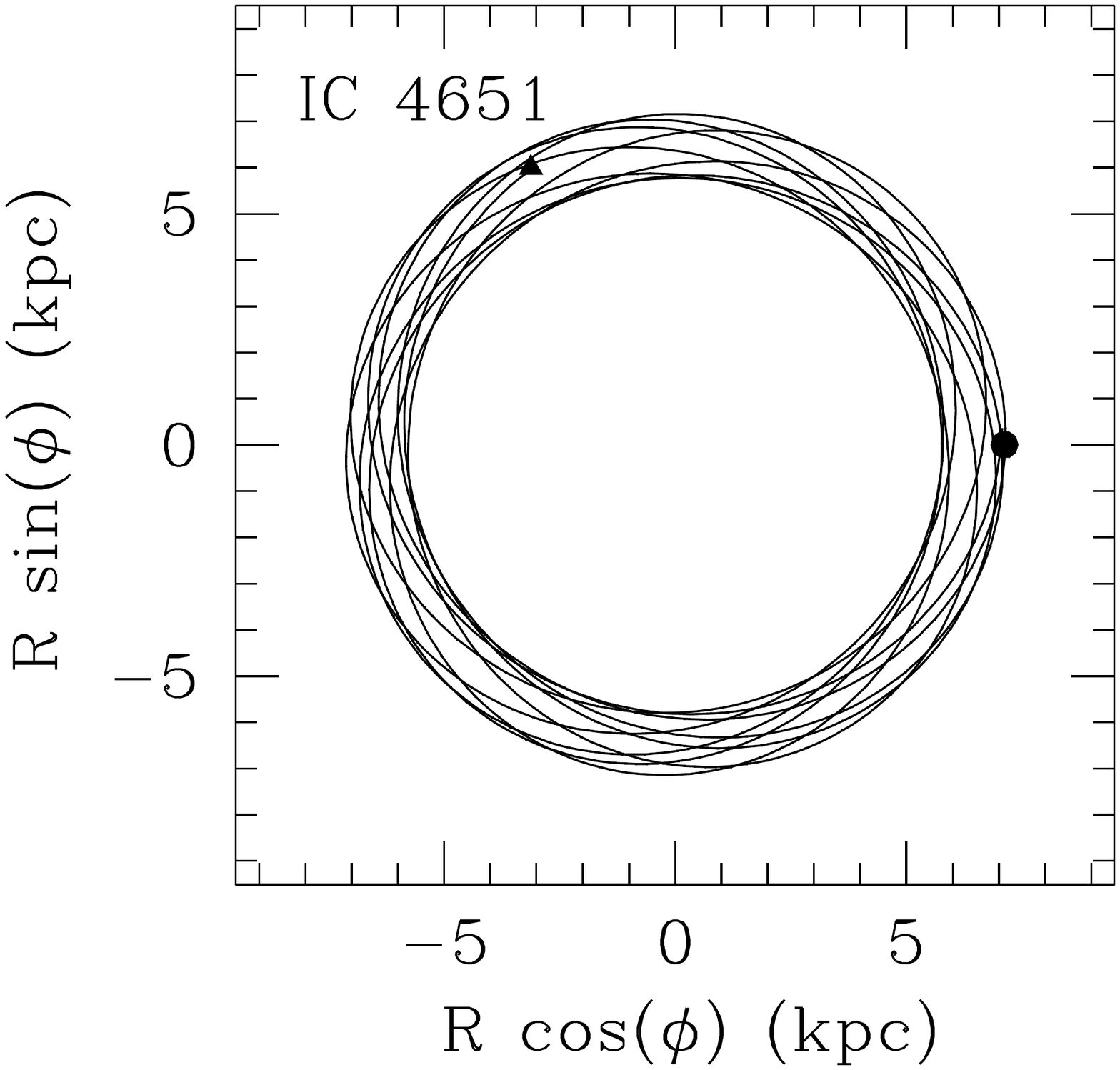}
    \caption{
     Galactic orbits of the clusters NGC 6067, NGC 2506 and IC 4651
     determined with Galactic potential model described in text. 
     The continuous
     line represents orbit of cluster in a time interval of age of cluster.
     For NGC 6067 the dotted line represents cluster orbit for a time 
     interval of 160 Myr. The left panel shows side picture and
     right panel shows top view of the orbit. The filled triangle and
     filled circle denote birth and present day position of cluster
     in the Galaxy. 
     }
    \label{6nm}
  \end{figure}

% Orbits
%---------------------------------------------------------------------------

\section{Orbits of the clusters} \label{sec:or}

\begin{table*}
   \centering
   \caption{Data used to derive orbits of the clusters. All
    parameters are calculated in present study.
   }
   \begin{tabular}{rccclrccc}
   \hline\hline
   Cluster   & $\alpha_{2000}$  & $\delta_{2000} $ &  Parallax & $ d_{\odot}$ &  Age & $v_{r}$ & $ \mu_{\alpha} cos{\delta}$ & $ \mu_{\delta}$   \\
   &  (deg) & (deg) & (mas) & (kpc) & (Myr) & (km/sec) & (mas/yr) & (mas/yr)   \\
  \hline
   NGC 6067 & 16:13:08.92  & -54:11:59  & $ 0.33 \pm 0.21 $ & $3.01 \pm 0.87$ & 66 &  $-38.70 \pm 0.95 $  &  $-1.90 \pm 0.01$ & $ -2.57 \pm 0.01 $    \\
NGC 2506 & 08:00:07.87 & -10:46:33  & $ 0.25 \pm 0.10 $ & $3.88 \pm 0.42$ & 2090 &  $ +84.64 \pm 1.06 $ & $ -2.57 \pm 0.01 $ & $ 3.92 \pm 0.01 $     \\
   IC 4651  & 17:24:48.44  & -49:56:07  & $ 1.00 \pm 0.07 $ & $1.00 \pm 0.08$  & 1590 &  $-30.18 \pm 1.07 $  & $ -2.41 \pm 0.01 $ & $ -5.05 \pm 0.02 $        \\
  \hline
  \end{tabular}
  \label{vinp}
  \end{table*}

\begin{table*}
   \centering
   \caption{Position and velocity components in Galactocentric
    coordinate system. Here $R$ is the Galactocentric distance, $Z$
   is vertical distance from the Galactic disc, $U$ $V$ $W$ are radial,
   tangential and vertical components of velocity respectively, $\phi$
   is the position angle relative to the sun's direction.
   }
   \begin{tabular}{lrrrcrccc}
   \hline\hline
   Cluster   & $R$ &  $Z$ &  $U$  & $V$  & $W$ & $\phi$   \\
   & (kpc) & (kpc) & (km/sec) &  (km/sec) & (km/sec) & (radians)    \\
  \hline
   NGC 6067 & 5.61 & -0.09 & $15.66 \pm 15.78 $  & $-220.82 \pm 17.00 $ &  $-0.35 \pm 0.20$  &  0.27    \\
   NGC 2506 &  10.84 & 0.69 & $ -43.99 \pm 0.58 $ & $ -248.08 \pm 12.13 $ & $ -17.27 \pm 0.30  $ & 0.28      \\
   IC 4651  & 7.03 & -0.12 & $ -16.89 \pm 1.40 $ & $ -223.62 \pm 1.18 $ & $ -7.55 \pm 0.20 $ & 0.05     \\
\hline
  \end{tabular}
  \label{inp}
  \end{table*}

\subsection{Initial conditions and Galactic model} \label{inic}

We derive orbits and orbital parameters of the
clusters under study, using Galactic potential models 
given by \citet{1975PASJ...27..533M} and \citet{1999MNRAS.310..645W}.
Input parameters used to calculate orbits
are listed in Table \ref{vinp}. Central coordinates $\alpha$ and $\delta$
of the clusters are taken from section
\ref{sec:cen}.
Mean 
proper motion ($\mu_{\alpha}cos\delta$, $\mu_{\delta}$) 
and heliocentric distance ($d_{\odot}$) for the clusters
are taken from section \ref{sec:kin}.
The age of the
clusters is considered as derived in section \ref{sec:iso}.

Since clusters are orbiting around the Galactic
centre, we cannot use position and velocity vectors in the equatorial system. 
Therefore,
we converted them into Galacto-centric
cylindrical coordinate system using the transformation matrix
given in \citet{1987AJ.....93..864J}. In this system $(r,\phi,z)$ denotes
the position of an object in Galaxy, where $r$ is a distance from Galactic centre, $\phi$ is
angle relative to Sun's position in Galactic plane and $z$ is the distance from
Galactic plane.

We adopted the right-hand coordinate system
to transform equatorial velocity components into Galactic-space
velocity components ($U,V,W$), where $U$, $V$ and $W$ are radial,
tangential and vertical velocities respectively.
In this system, $U$ is taken positive towards Galactic-centre, $V$ is
along the direction of Galactic rotation and $W$
is towards Galactic north pole.
The Galactic centre is taken at ($17^{h}:45^{m}:32^{s}.224, -28^{\circ}:56^{\prime}:10^{\prime\prime}$)
and North-Galactic pole is at
($12^{h}:51^{m}:26^{s}.282, 27^{\circ}:7^{\prime}:42^{\prime\prime}.01$) 
\citep{2004ApJ...616..872R}.
The velocity components of clusters for Local Standard of Rest (LSR)
are calculated using space-velocity components of Sun given by \citet{1968gaas.book.....M}
as ($-10.4, +237, +7.3$) km/s.
After that velocity are corrected for Galactic standard of rest (GSR)
using position
coordinates of Sun as ($8,0,0.02$) and velocity of LSR as 220 km/s.
Transformed parameters in Galacto-centric coordinate system are listed
in Table \ref{inp}.

Radial velocity data for the clusters NGC 6067, NGC 2506 and IC 4651
are available for 26, 37 and 67 cluster members respectively in Gaia DR2 catalogue.  
Average radial velocities of the clusters are calculated by
taking the mean
for all stars 
and are found
as $-38.70 \pm 0.95$, $84.64 \pm 1.06$ and $-30.18 \pm 1.07$ km/s
for the clusters NGC 6067, NGC 2506 and IC 4651 respectively.

For Galactic potentials, we adopted an approach
given by \citet{1991RMxAA..22..255A}. According to their model, the mass of Galaxy is
described by three components: spherical central bulge,
massive spherical halo and disc. In the present model, we used
axis-symmetrical, time-independent and analytic potential models
for all three components. Also, all potentials and their spatial derivatives are
continuous everywhere.
We used \citet{1975PASJ...27..533M} potentials for bulge and disc regions,
and \citet{1999MNRAS.310..645W} potential
for halo region.
These potentials are given as    \\

$ \Phi_{b}(r,z) = -\frac{M_{b}}{\sqrt{r^{2} + b_{b}^{2}}} $   \\

$ \Phi_{d}(r,z) = - \frac{M_{d}}{\sqrt{r^{2} + (a_{d} + \sqrt{z^{2} + b_{d}^{2}})^{2}}}  $ \\

$ \Phi_{h}(r,z) = - \frac{M_{h}}{a_{h}} ln(\frac{\sqrt{r^{2} + a_{h}^{2}} + a_{h}}{r}) $    \\

Where  $ \Phi_{b} $ , $ \Phi_{d} $ and $ \Phi_{h} $ are the potentials
of the central bulge, disc and halo of Galaxy respectively. $r$ and $z$ are
the distances of objects from Galactic centre and Galactic disc
respectively.
Values of the constants used in these equations are 
taken from \citet{1975PASJ...27..533M} and \citet{1999MNRAS.310..645W}.

\subsection{Orbits calculation}

Using the initial conditions and Galactic potential model described
in section \ref{inic}, we calculated the orbits of the clusters under study.
In orbit determination, we estimated 
the radial and vertical components of gravitational force,
by differentiating total gravitational potentials with respect to
$r$ and $z$. The second-order derivatives of the
gravitational force describe the motion of the clusters.
The second-order derivatives
are integrated backwards
in time, which is equal to the age of clusters.
NGC 6067 is a young cluster and has not yet completed
one revolution around the Galactic centre. To 
calculate the eccentricity of its orbit, we integrate orbit up-to a
time interval of 160 Myr.
Since potentials used are axis-symmetric,
energy and $z$ component of angular momentum are conserved throughout
the orbits.

Fig. \ref{6nm} shows orbits
of the clusters NGC 6067, NGC 2506 and IC 4651. 
In left panels, the motion of clusters is described in terms of distance
from Galactic centre and Galactic plane, which shows 
a two-dimensional side view of the orbits. 
In right panels, cluster motion is described in terms of
$x$ and $y$ components of Galactocentric distance, which shows
the top view of orbits. 

We also calculated the orbital parameters for the clusters and are listed
in Table \ref{orpara}. Here $e$ is the eccentricity, $R_{a}$ is
apogalactic distance, $R_{p}$ is perigalactic distance, $Z_{max}$
is maximum distance travelled by cluster from Galactic disc,
$E$ is the average energy of orbits,
$J_{z}$ is $z$ component of angular momentum and $T$ is the time period
of the clusters in the orbits.

\begin{table*}
   \centering
   \caption{Orbital parameters of clusters calculated
    using Galactic potential model described in the text.
   }
   \begin{tabular}{lcccccccc}
   \hline\hline
   Cluster  & $e$  & $R_{a}$  & $R_{p}$ & $Z_{max}$ & $E$ & $J_{z}$ & $T$   \\
           &    & (kpc) & (kpc) & (kpc) &  $(100 km/sec)^{2}$ & (100 kpc km/s) & (Myr) \\
   \hline\hline
   NGC 6067 & $ 0.008 \pm 0.294$ & $ 5.66 \pm 1.64$ &  $ 5.57 \pm 1.61$ & $ 0.09 \pm 0.03$ &  $ -17.32 \pm 0.23$ & $ -12.39 \pm 3.71$ & $ 156 \pm 47$ \\
   NGC 2506  & $ 0.002 \pm 0.109$ & $13.06 \pm 1.41$ & $ 12.99 \pm 1.41$  & $ 0.93 \pm 0.07$  & $ -12.57 \pm 0.12$  & $ -26.88 \pm 3.19$ & $ 273 \pm 32$  \\
   IC 4651  & $ 0.001 \pm 0.080$  &  $ 7.15 \pm 0.57$ & $ 7.16 \pm 0.57$ & $ 0.15 \pm 0.01$ & $ -15.84 \pm 0.02$ & $ -15.82 \pm 1.27$ & $ 196 \pm 16$  \\
  \hline
  \end{tabular}
  \label{orpara}
  \end{table*}

The orbits of all three clusters under study follow a boxy pattern.
The eccentricity of all the clusters is nearly zero. 
Hence they
trace a circular path around the
Galactic centre.
From these orbits 
we determine, the birth and present-day position
of clusters in the Galaxy and are denoted by the filled triangle and 
the filled circle in Fig. \ref{6nm}.

NGC 6067 being a young cluster
has not yet completed even one revolution around the Galactic center
as seen in Fig. \ref{6nm}.
The motion of this cluster is confined inside the solar circle within
a box of $ 4.5 < R_{gc} \leq 6.7 $ kpc and is
oscillating near the Galactic plane.
Similar behaviour of the orbit is also seen in Fig. \ref{6nm} for the
cluster IC 4651. This cluster is also orbiting
inside the solar circle within a box of $ 5.8 < R_{gc} < 7.3 $ kpc
and not very far from Galactic plane ($Z_{max}$=0.15 kpc).

NGC 2506 is the oldest (2.09 Gyr) cluster in our study. The eccentricity
of the orbit of this cluster is $\sim 0.00$ which indicates
a circular orbit around the Galactic centre. 
Orbit is confined in a box of $9.8 < R_{gc} \leq 13.2$ kpc
and therefore this cluster is not interacting with the inner region
of the Galaxy. The similar results are also shown by
\citet{1994A&A...288..751C} for the cluster NGC 188.
\citet{2014MNRAS.445.1048W}
studied the effect of eccentricity on the evolution of clusters and
concluded that clusters having eccentric
orbits evolve faster than the clusters having circular orbits.
So present analysis of the cluster NGC 2506 indicates that this 
cluster is evolving slowly. Being at a larger distance from Galactic
centre, the effect of Galactic tidal forces
on NGC 2506 will be less as compared to IC 4651. Because of this,
in spite of having similar ages,
NGC 2506 is more bound and compact than IC 4651.

Orbital parameters determined in the present analysis are very much similar
to the parameters found by \citet{2009MNRAS.399.2146W},
except that their orbits are more eccentric than what we found in the present analysis.
According to \citet{2006A&A...451..515M} the Galactic disk is warped and flared after
$R_{gc} \sim 15$ kpc. NGC 2506 traces the Galaxy
up-to a maximum distance of 13 kpc so it is orbiting near the
warped and flared disk but its present-day position ($R_{gc}=10.84$ kpc)
is slightly distant from the warped and flared disk.

%  Analysis
%----------------------------------------------------------------------

\section{Basic Parameters of the Clusters} \label{sec:ana}

\subsection{Cluster centre and radius} \label{sec:cen}

Though open clusters are very sparse and loosely bound systems but at 
centre, maximum stellar density (number of stars per unit area)
is observed. One can roughly estimate cluster centre by simply looking at
the cluster. To determine cluster centre precisely, we calculated
an average $X$ and $Y$ positions of stars iteratively
within 2000 pixels from eye estimated centre, until they
converge to a constant value. We expect an error
of a few tens of pixels in locating the cluster centre. In this way
the central coordinates obtained are (3994, 4406),
(4342,4246) and (4070, 4019) pixels for
NGC 6067, NGC 2506 and IC 4651 respectively.
The corresponding central coordinates of the clusters in the equatorial coordinate system are 
listed in Table \ref{vinp}. Present estimated values are 
similar to the values listed in Table \ref{bpara} which are taken from the WEBDA
database.

\begin{figure}
    \centering
    \includegraphics[width=8cm,height=8cm]{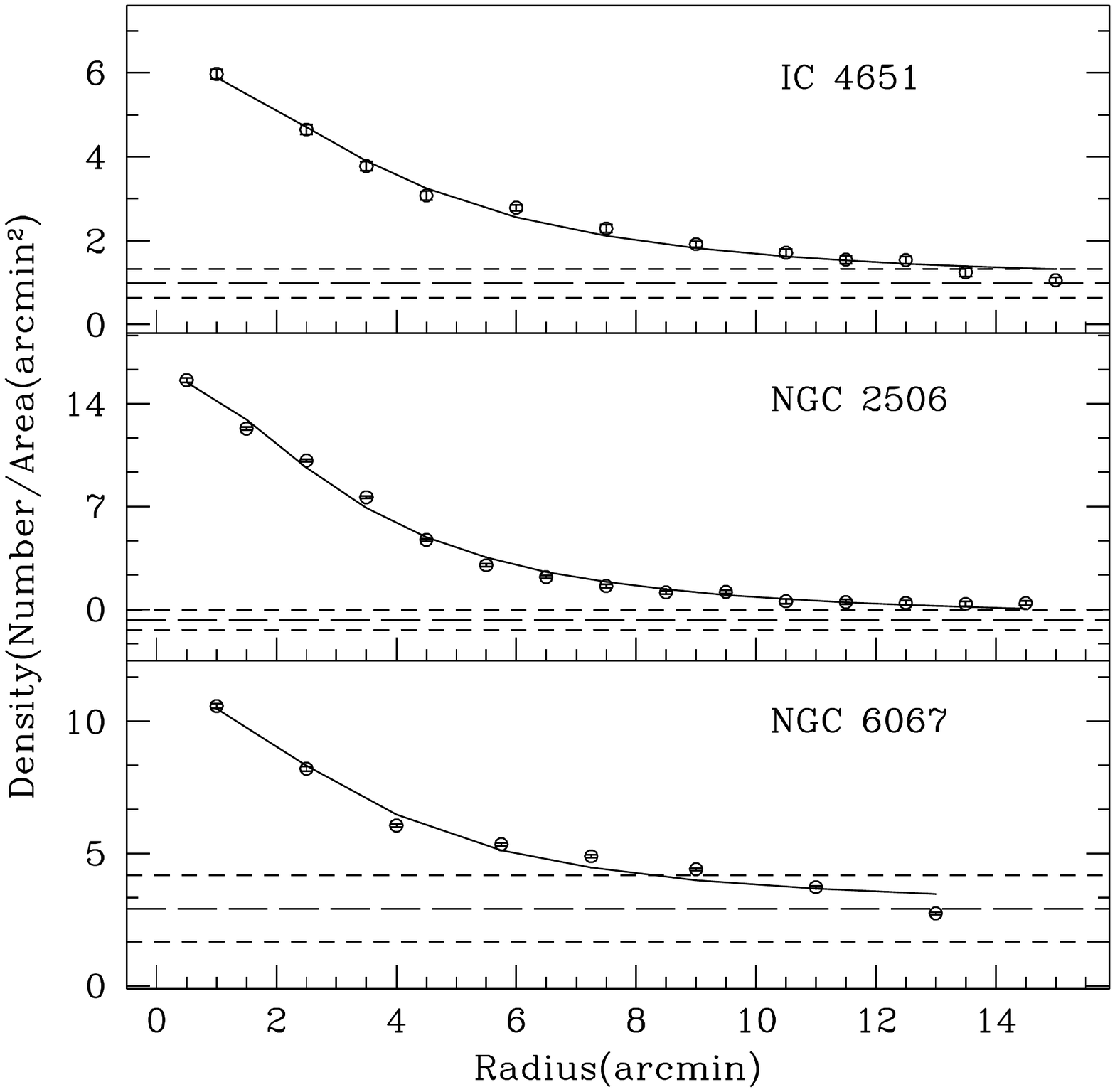}
    \caption{The radius is plotted against the number of stars
    per unit area for the clusters under study. The continuous 
    curve represents King's profile, which is 
    fitted on the data points. Error bars are plotted for each
     point. The horizontal long dashed middle lines represent the 
    background stellar density level while the short dashed 
    lines represent the 1$-\sigma$ upper and lower limits in background stellar density.
       }
    \label{rdp_f}
  \end{figure}

The wide-field images studied here for clusters under study
cover a large area which is sufficient enough 
to derive the true cluster extent and radial surface density profile (RDP) of the
clusters.
We divided the entire cluster region into several concentric
rings of thickness $1^{\prime}$ around
the cluster center. 
The stellar density 
of each ring is calculated by dividing the number of stars
by area of that ring.
Stellar densities as a function 
of cluster radius 
are shown in Fig. \ref{rdp_f} for all three clusters.
We fitted \citet{1962AJ.....67..471K}
profile on the radial density profiles which is
shown by continuous curve in Fig. \ref{rdp_f}. 
The King profile is given by:   \\

$f(r) = {f_{b} + {f_{0} \over {1+(r/r_{c})^{2}}}}$   \\ 

where $f_{0}$ is the central density, $r_{c}$ is core radius and $f_{b}$
is the background density. Core radius is the
distance from the cluster center at which the stellar density becomes
half of the central density.
The fitting gives 
different parameters of clusters which are listed
in Table \ref{sp}. The radius of clusters
are determined as $10^{\prime}$, $12^{\prime}$ and $11^{\prime}$ 
for the clusters NGC 6067, NGC 2506 and
IC 4651 respectively. The stellar central densities show
that NGC 2506 is a dense cluster in comparison to NGC 6067 and IC 4651 clusters.

\begin{table}
   \centering
    \small
   \caption{Structural parameters of the clusters, calculated using the
   radial density profile. Radius and $r_{c}$ are in arcmin while $f_{0}$ and $f_{b}$ are
   in the unit of stars per $arcmin^{2}$.}
   \begin{tabular}{lcrrr}
   \hline\hline
  Cluster & Radius  & $f_{0}$ & $r_{c}$ & $f_{b}$   \\
  \hline
  NGC 6067 & 10  &  $08.18 \pm 0.60$ & $3.51 \pm 0.55$ &  $2.92 \pm 0.42$  \\
  NGC 2506 & 12  &  $16.58 \pm 0.38$ & $3.23 \pm 0.15$  &  $-0.73 \pm 0.22$     \\
  IC 4651  & 11  &  $5.22 \pm 0.19$ & $3.95 \pm 0.28$  &  $0.98 \pm 0.11$    \\
  \hline
  \end{tabular}
  \label{sp}
  \end{table}

% Isochrones
%---------------------------------------------------------------------

\subsection{Age and Distance of the clusters using isochrone fitting} \label{sec:iso}

\begin{figure*}
    \centering
    \includegraphics[width=8cm]{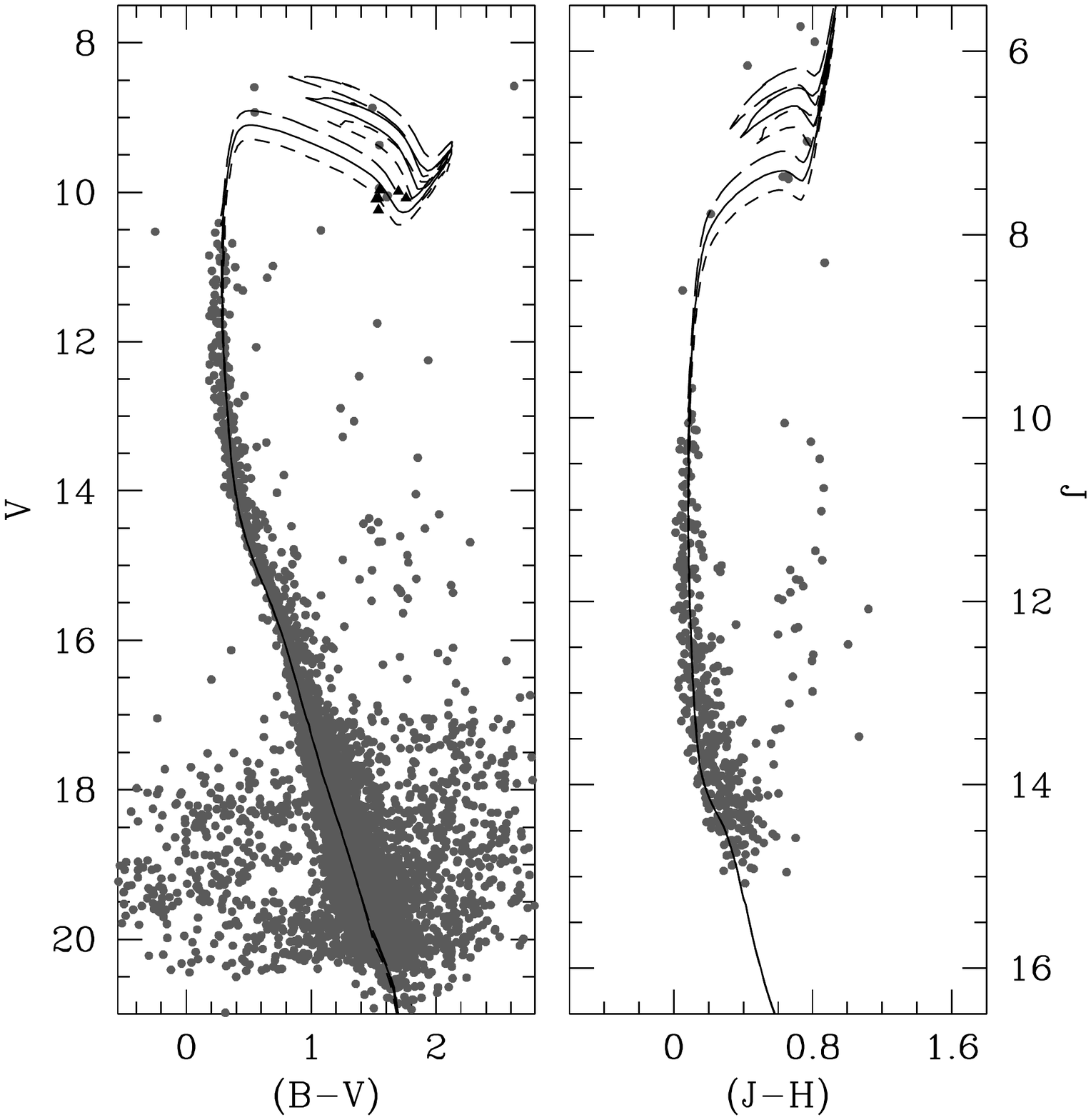}
    \includegraphics[width=8cm]{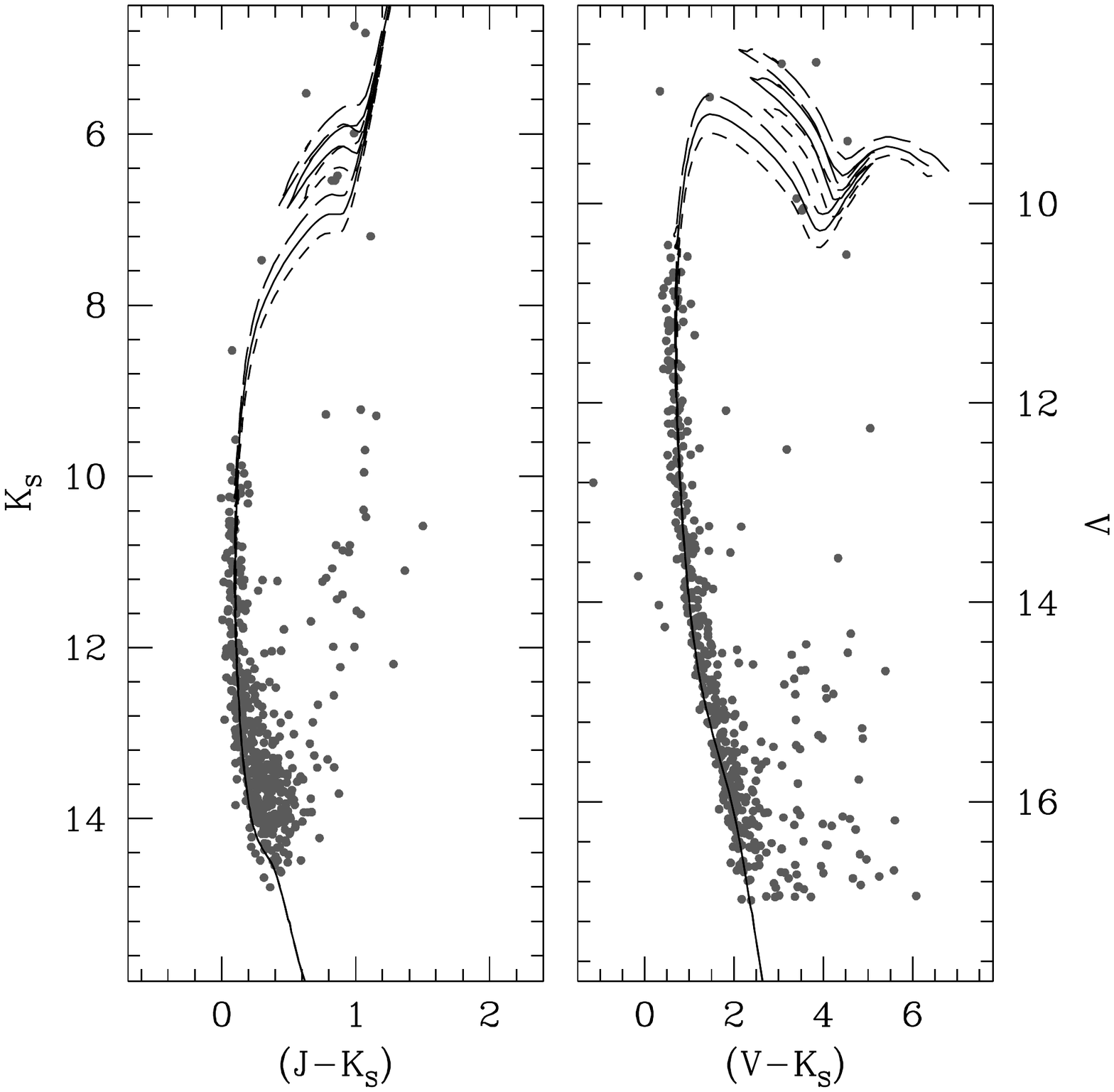}
    \caption{$(V,B-V)$, $(J,J-H)$, $(K_{s},J-K_{s})$ and $(V,V-K_{s})$
    colour-magnitude diagrams for the cluster NGC 6067 using
    most probable cluster members. \citet{2017ApJ...835...77M} isochrones
    of log(age) 7.77, 7.82 and 7.87 are fitted on cluster sequence.
    Filled triangles in $(V,B-V)$ plot are stars taken
    from \citet{2007ApJ...671.1640A}.}
    \label{iso6067}
  \end{figure*}

To determine age, metallicity and reddening of clusters, we used 
theoretical evolutionary isochrones given by \citet{2017ApJ...835...77M}
which are derived from the stellar evolutionary tracks computed with
PARSEC \citep{2012MNRAS.427..127B} and COLIBRI 
\citep{2013MNRAS.434..488M} codes. 
They include thermally pulsating asymptotic giant branch phase
for solar metallicity 
$Z_{\odot}=0.01524$ and $Y_{\odot}=0.2485$ \citep{2011SoPh..268..255C} with 
mixing length and [$\alpha/$Fe] taken as 1.74 and zero respectively. 
For colour-magnitude diagrams, we considered only those stars which have 
a high probability to be cluster members based on the selection done in section \ref{sec:vpds}.
2MASS data for cluster member stars are used to revisit the cluster
parameters. The uncertainties in the age and distance reflect the range in 
these parameters that allows a reasonable fit to the cluster colour-magnitude
diagram.
The fitting of isochrones
for the three clusters under study are discussed as follows;

\begin{figure*}
    \centering
    \includegraphics[width=5cm,height=5cm]{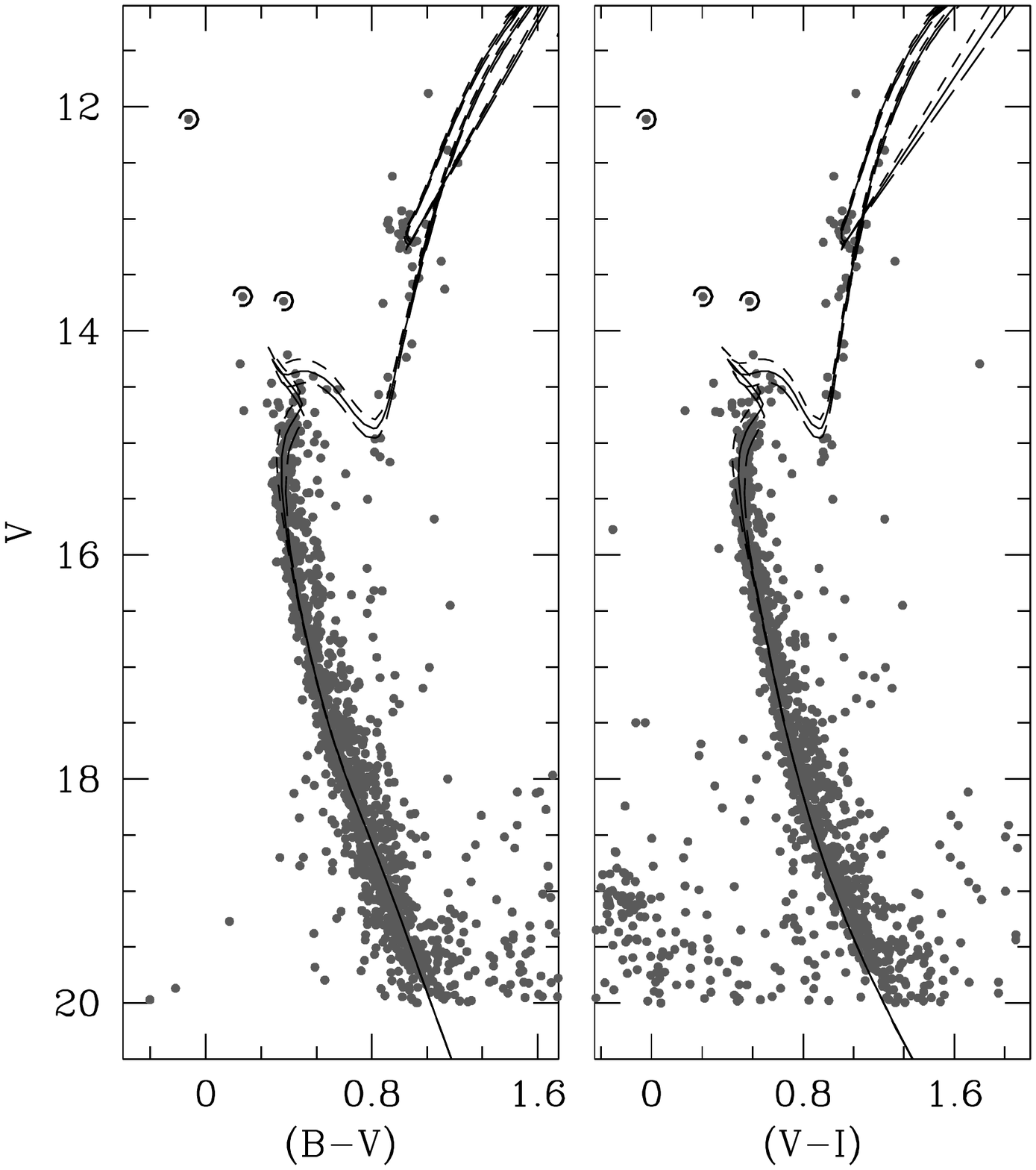}
    \includegraphics[width=5cm,height=5cm]{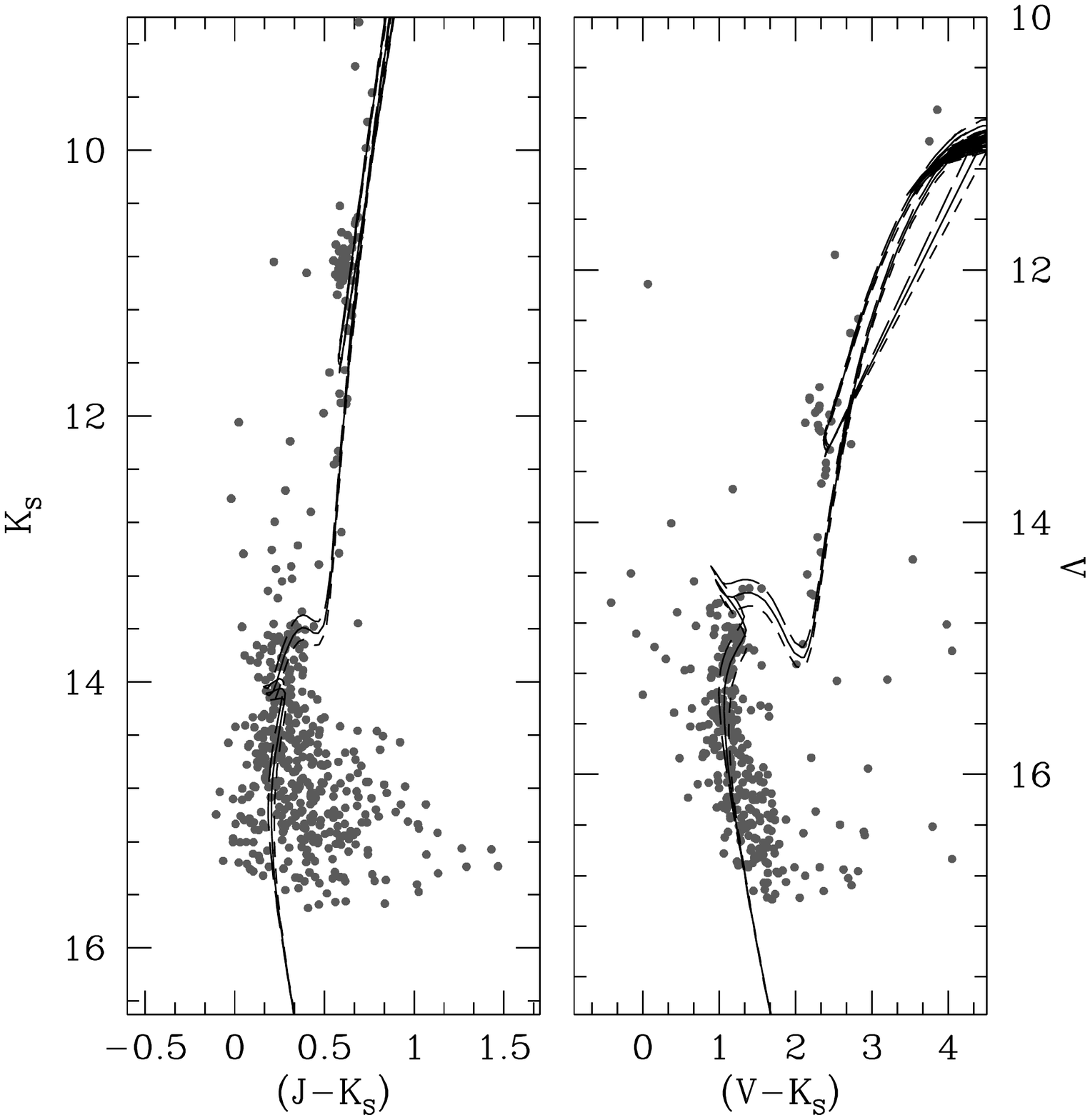}
    \includegraphics[width=5cm,height=5cm]{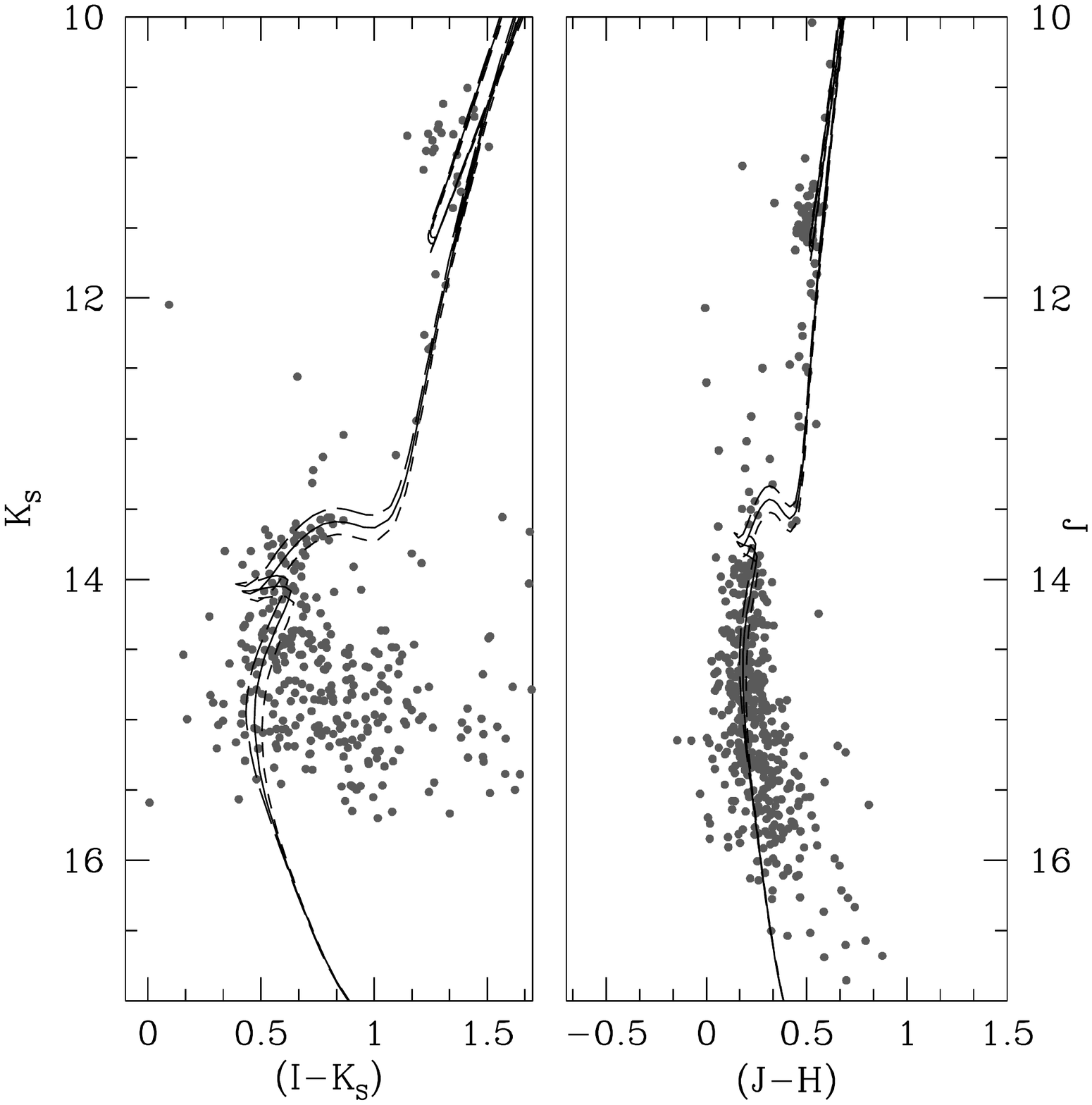}
    \caption{$(V,B-V)$, $(V,V-I)$), $(K_{s},J-K_{s})$, 
      $(V,V-K_{s})$, $(K_{s},I-K_{s})$ and $(J,J-H)$
     colour-magnitude diagrams for the cluster NGC 2506
    with the best fitted \citet{2017ApJ...835...77M} isochrones 
    of log(age) 9.29, 9.32 and 9.35 on 
    the cluster sequence. Points within Circles in optical colour-magnitude
    diagrams denote blue straggler stars 
    identified in this analysis.}
    \label{iso2506}
  \end{figure*}

{\it NGC 6067}: We plotted colour-magnitude diagrams
 for the cluster in $(V,B-V)$, $(J,J-H)$,
$(K_{s},J-K_{s})$ and $(V,V-K_{s})$ 
planes as shown in Fig. \ref{iso6067}.  
We performed several iterations by taking isochrones 
of different metallicity and age. 
The isochrones of $Z= 0.02$ and age ( $66 \pm 0.8$ ) Myr
is satisfactorily fitted. The distance modulus is
derived as $13.0 \pm 0.2$ mag, which gives heliocentric distance
as $2.09 \pm 0.2$ kpc. This distance is comparable to the distance calculated by the parallax angle.
Distance modulus derived by Alanso-Santiago et al. (2017) is much different than the 
present estimate. Alanso-Santiago et al. (2017) used photographic data for distance modulus 
estimation. The present estimate of distance modulus in more reliable because it is based on parallax 
as well as CCD optical and near-IR data. Isochrone fitting also gives the values of colour excess 
$E(B-V)$, $E(J-H)$, $E(J-K_{s})$ and
$E(V-K_{s})$ as 0.45, 0.15, 0.22 and 1.2 mag. Using these colour excess
values we derived the colour excess ratios as $E(J-H)/E(B-V) = 0.33$, 
$E(J-K_{s})/E(B-V) = 0.488$
and $E(J-K_{s} )/E(J-H) = 1.5$. 
These values are in good agreement with the relations given 
by \citet{2002A&A...381..219D, 2012A&A...539A.125D}.

{\it NGC 2506}: We plotted six colour-magnitude diagrams
($(V,B-V)$, $(V,V-I)$), $(K_{s},J-K_{s})$, $(V,V-K_{s})$, $(K_{s},I-K_{s})$ 
and $(J,J-H)$ as shown in Fig. \ref{iso2506}.
The best-fitted isochrone in all colour-magnitude diagrams is found to be of
$Z = 0.007$ and age $2.09 \pm 0.14$ Gyr.
The distance modulus for
the cluster is calculated as $12.7 \pm 0.2$ mag. 
The derived heliocentric distance 
is $3.21 \pm 0.30$ kpc. The distance derived from isochrone
fitting is in good agreement with that of the parallax
angle method discussed in section \ref{sec:parallax}.
Values of colour-excess $E(B-V)$, $E(V-I)$, $E(V-K)$, $E(J-H)$, $E(J-K_{s})$ and
$E(I-K_{s})$ are estimated as 0.06, 0.07, 0.28, 0.02, 0.03 and 0.08 mag.

Three blue straggler stars are also identified in the present analysis. These 
are shown with the open circles in optical colour-magnitude diagrams
of the Fig. \ref{iso2506}.
Motions of the three blue straggler stars
have consistency with the cluster motion and have the same
distances as the other cluster members, hence they are a member of 
this cluster. Their spatial location in the cluster shows that they are 
centrally concentrated.

\begin{figure*}
    \centering
    \includegraphics[width=5cm]{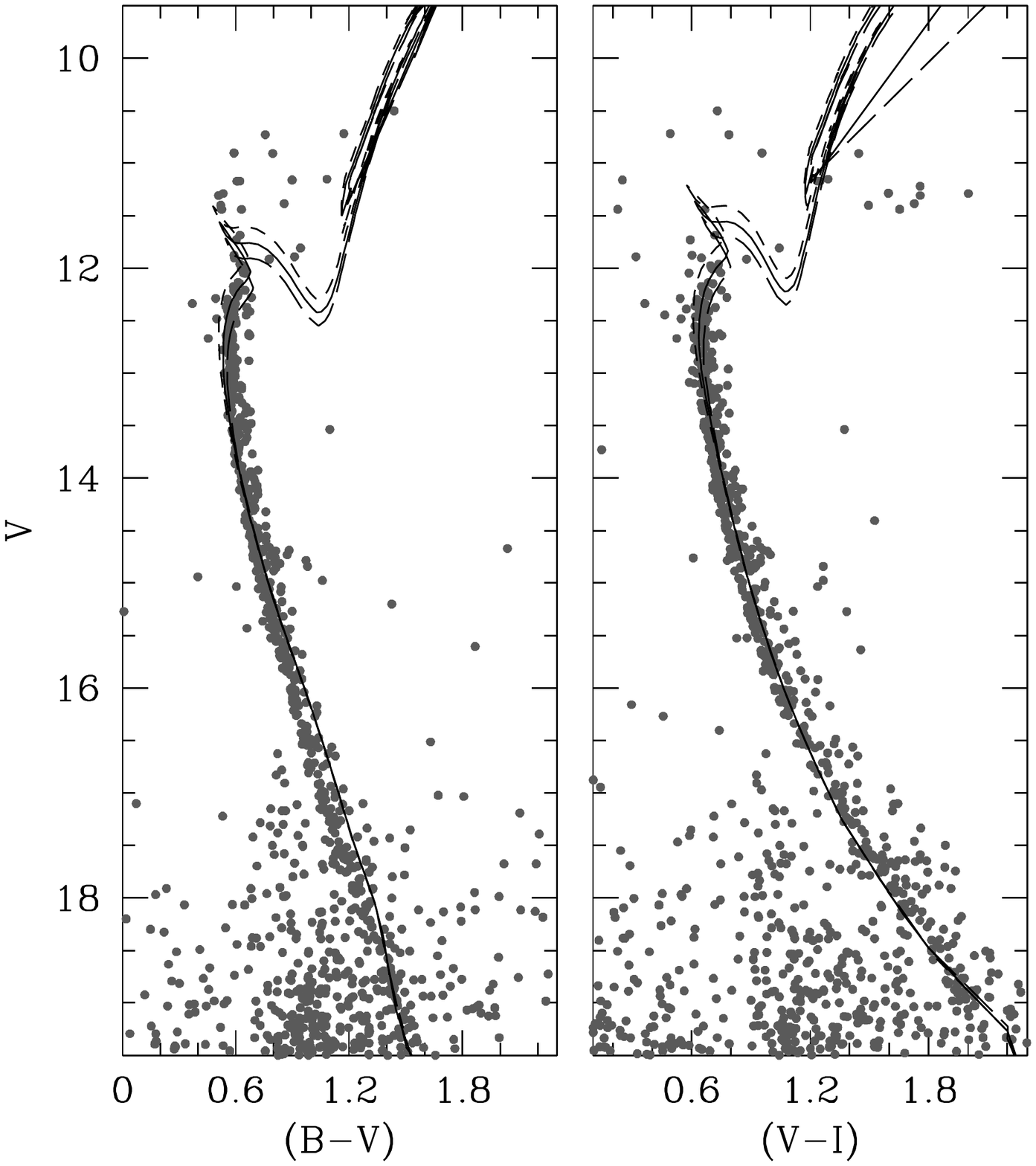}
    \includegraphics[width=5cm]{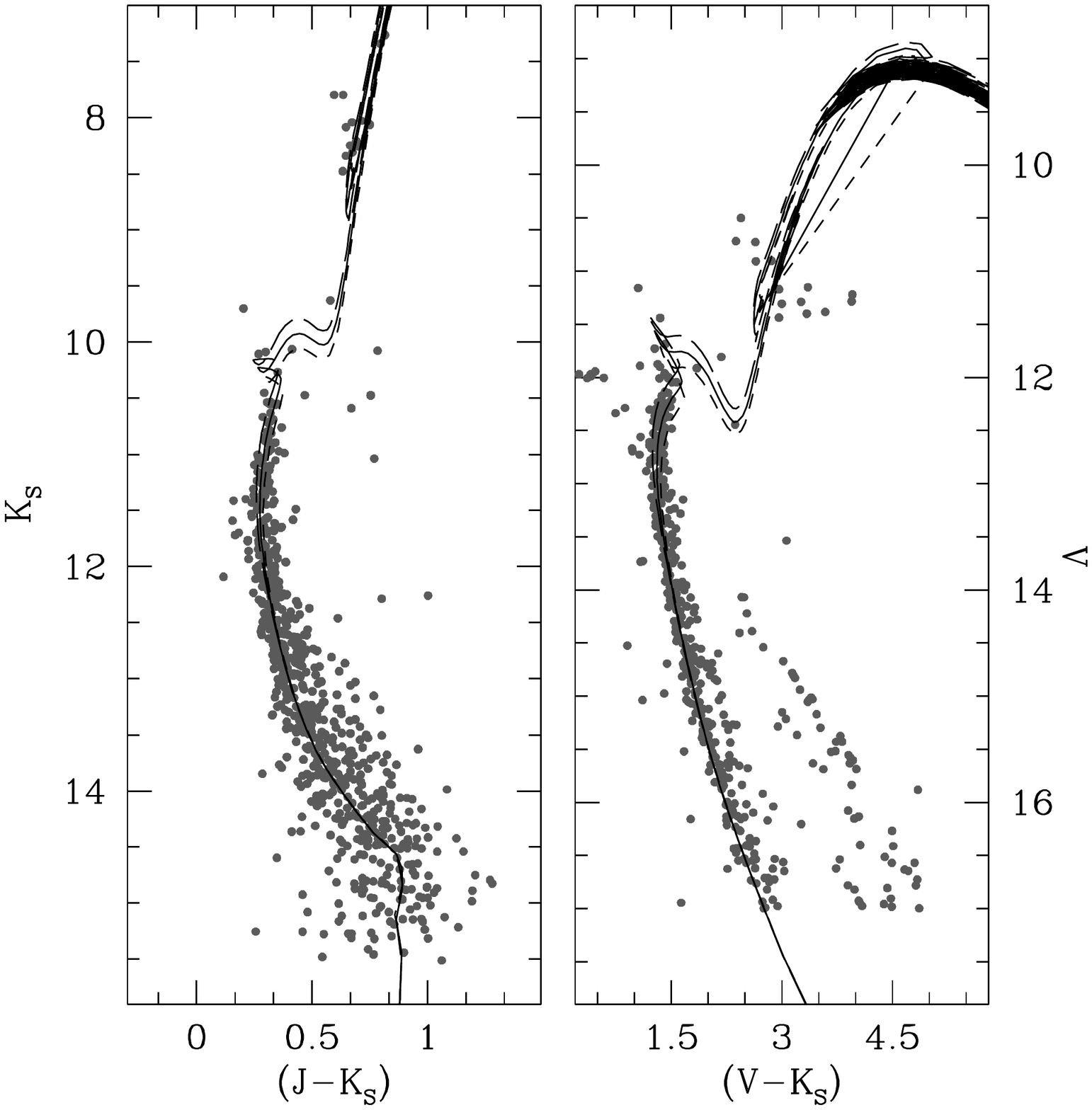}
    \includegraphics[width=5cm]{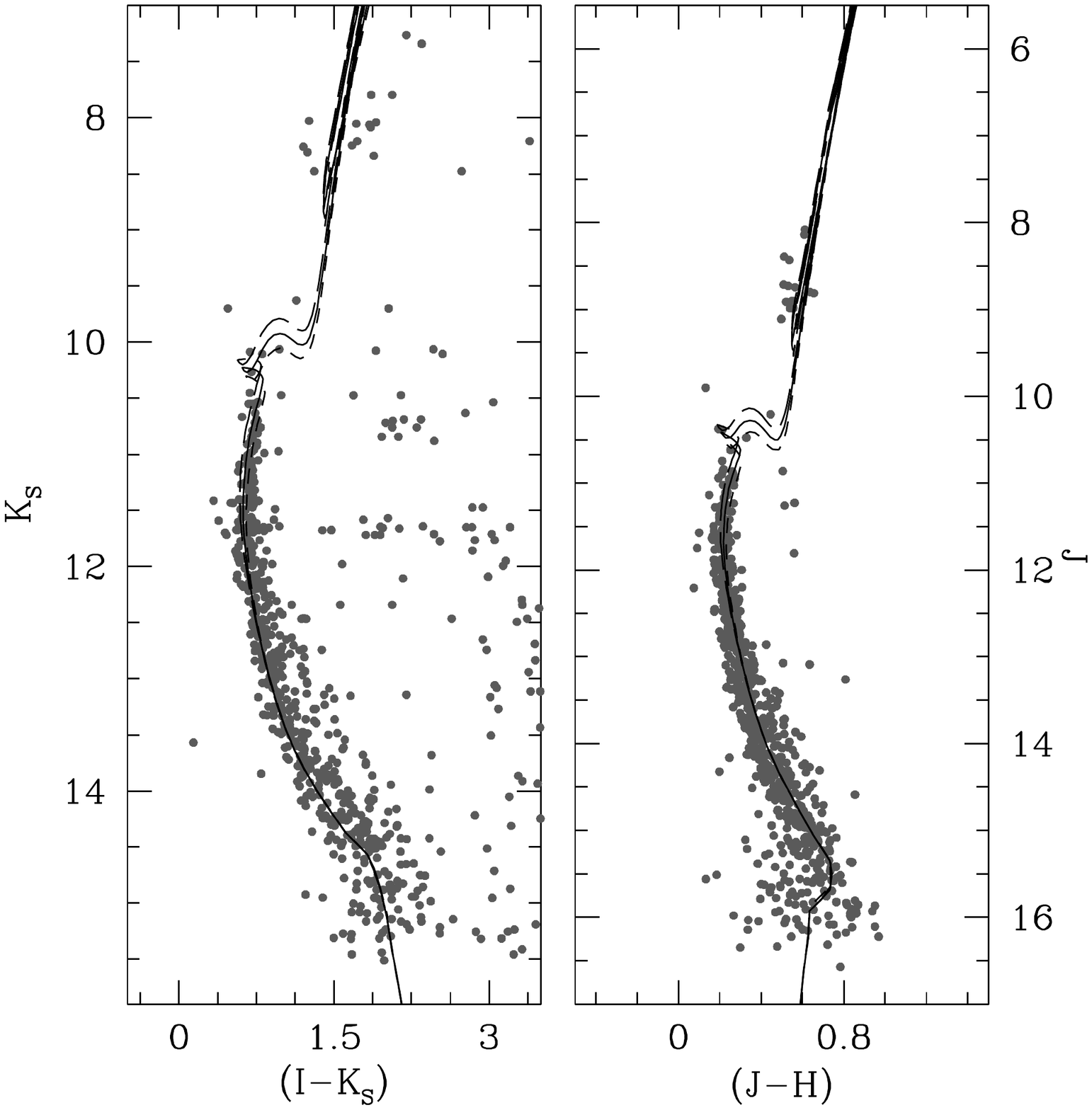}
    \caption{$(V,B-V)$, $(V,V-I)$, $(K_{s},J-K_{s})$, $(V,V-K_{s})$,
    $(K_{s},I-K_{s}$ and $(J,J-H)$ colour-magnitude diagrams
    for the cluster IC 4651. The best fitted isochrones of log(age)
    9.20, 9.16 and 9.24 are indicated by continuous, short dashed and
    long dashed lines respectively.
    }
    \label{iso4651}
  \end{figure*}

{\it IC 4651}: Fig. \ref{iso4651} shows $(V,B-V)$, $(V,V-I)$,
$(K_{s},J-K_{s})$, $(V,V-K_{s})$, $(K_{s},I-K_{s}$
and $(J,J-H)$ colour-magnitude diagrams
using the stars selected in section \ref{sec:vpds}.
The best-fitted isochrone
for all the colour-magnitude diagrams is found to be of $Z \sim 0.019$ and  
age $1.59 \pm 0.14$ Gyr.
Distance modulus  
for the cluster is derived as $10.4 \pm 0.2$ mag and 
corresponding heliocentric distance 
is as $0.96 \pm 0.09$ kpc. The value of $E(B-V)$, 
$E(V-I)$, $E(V-K_{s})$, $E(J-H)$, $E(J-K_{s})$ and
$E(I-K_{s})$ are determined as 0.16, 0.2, 0.44, 0.04, 0.08 and 0.19 mag.
In this cluster the stars brighter than $V \sim 12$ mag are saturated in
our photometry in all filters.
In $(V,V-I)$ colour-magnitude diagram of the cluster 
a binary sequence is also visible parallel to the main-sequence.

A comparison of parameters obtained in our analysis with the values
available in the
literature is presented in Table \ref{ctab}. This table shows that 
presently determined parameters are
in good agreement with previous studies. 
\citet{2018MNRAS.478..651G} derived a three-dimensional
interstellar dust reddening map covering sky north of $ \delta \sim -30^{\circ}$ using photometry of 
800 million stars from Pan-STARRS 1 and 2MASS.
For NGC 2506, their estimated value of 
$E(B-V)$ is $0.07 \pm 0.02$ 
which is in good agreement with our determined value. 
NGC 6067 and IC 4651 do not lie in the region covered by the reddening map.

\begin{table*}
   \centering
   \small
   \caption{A comparison of different parameters of the clusters with
   literature. DM is the distance modulus of the clusters.}
   \begin{tabular}{lcrll}
   \hline\hline
  Author & Radius  & Age & DM  &  $E(B-V)$   \\
        & (arcmin) &  (Myr)  & (mag) &  (mag)  \\
  \hline
  &{\bf NGC 6067}&&&   \\
  \citet{1962MNRAS.124..445T} & - & 75 to 150  &  12.5 &    0.33   \\
  \citet{1985MNRAS.214...45W}  & -  & -  & $ 11.05 \pm 0.1 $  &   0.35  \\
  \citet{2017MNRAS.469.1330A}  & $ 14.8 \pm 6.8 $  & $90 \pm 20$   & $11.25 \pm 0.15$ &  $ 0.35 \pm 0.04$   \\
  Present study & 10  & $ 66 \pm 8.0 $  &  $13.0 \pm 0.2 $ &  0.45   \\
 \hline
  &{\bf NGC 2506}&&&      \\
  \citet{1981ApJ...243..841M} & - & 3400  &  12.2  &   -  \\
  \citet{1997MNRAS.291..763M} & - & 1500 to 2200   & 12.5  &  - \\
  \citet{2012MNRAS.425.1567L} & - & $2310 \pm 160$ & $12.47 \pm 0.08$ &  $0.03 \pm 0.04$ \\
  \citet{2016AJ....152..192A} &  -  & $1850 \pm 50 $ & $ 12.75 \pm 0.1 $  & $ 0.058 \pm 0.001 $    \\
  Present study & 12  & $ 2090 \pm 140 $  & $12.70 \pm 0.20 $ &  0.06   \\
  \hline
  &{\bf IC 4651}&&&      \\
  \citet{2000AJ....119.2282A} &  -  & $ 1700 \pm 100 $ & 10.15  &   -    \\
   Meibom (2000) & - & - & $ 10.03 \pm 0.1 $  & -  \\
   Biazzo et al. (2007) & - & $ 1200 \pm 200 $ & - &  $ 0.12 \pm 0.02 $ \\
  Present study & 11  & $1590 \pm 140 $  & $10.40 \pm 0.20 $   & 0.16   \\
  \hline
  \end{tabular}
  \label{ctab}
  \end{table*}

% Completeness  and lf
%---------------------------------------------------------------------------

\section{Luminosity and mass function of the clusters} \label{sec:luminosity}

\subsection{Determination of photometric completeness} \label{comp}

To calculate the completeness level in our photometry, 
we conducted artificial star (AS) test.
For completeness determination in NGC 6067, $B$ and $V$ images are 
used whereas for NGC 2506 and IC 4651 $V$ and
$I$ images are used.
For AS test we added total $1.5 \times
10^{5}$ artificial stars in each image. 
These stars
must have the same geometric position
and different magnitudes in the two filters.

The AS test routine added stars one by one from the artificial star list.
After addition of a star, AS routine perform photometry on the image,
with the same method which was used for the real stars. 
The same procedure was performed for another star and so on.
To recover the stars in AS test, we considered the conditions 
explained in \citet{2012ApJ...754L..34M}. According to the conditions, an artificial star can
be considered as recovered  when the input and the output fluxes differ
by $< 0.75 $ mag and their positions by $<0.5 $ pixel. 

The completeness factor (CF) is calculated as the ratio of total 
recovered stars to the total number of added stars in a magnitude bin in both
the filters. 
We considered only those stars which are
recovered in both the filters.
In Table \ref{compt}, we listed the CF in different magnitude bins
for the clusters under study. This table shows that stars 
brighter than 14 mag has CF $\sim$ 99.99 $\%$ for all clusters.
 
\begin{table}
   \centering
   \caption{The photometric completeness of the data in each
   magnitude bin for the clusters under study.
   }
   \begin{tabular}{ccccccccc}
   \hline\hline
 $ V $ &  NGC 6067 & NGC 2506 & IC 4651   \\
  (Mag) &     &   &   \\
  \hline
   12 - 13 &  99.99  &  99.99  &  99.99    \\
   13 - 14 &  99.99  &  99.99  &  99.99    \\
   14 - 15 &  97.18  &  98.49  &  91.79    \\
   15 - 16 &  89.40  &  98.19  &  90.38    \\
   16 - 17 &  86.49  &  97.87  &  89.89    \\
   17 - 18 &  89.56  &  96.92  &  88.00    \\
   18 - 19 &  87.09  &  94.66  &  82.43    \\
   19 - 20 &  81.26  &  93.33  &  70.70    \\
   20 - 21 &  74.11  &  85.80  &  53.28    \\
  \hline
  \end{tabular}
  \label{compt}
  \end{table}

\subsection{Luminosity function} \label{subsec:l}

The number of stars in a unit magnitude
range is called luminosity function. To derive luminosity function, we need 
true cluster members, so it is important to apply the corrections of
non-members and the completeness.
For this analysis, we selected cluster members using proper motion data
as discussed in section \ref{sec:vpds}.
In spite of selecting members using proper motions and parallax angle,
some field stars are also seen in cluster colour-magnitude diagrams
shown in Fig. \ref{iso6067}, \ref{iso2506} and \ref{iso4651}.  
To overcome this problem, we applied photometric criteria to
select the cluster members as discussed in \citet{2008MNRAS.390..985Y}. 
In this method we make a blue and red envelope around the
cluster main-sequence. The stars inside this envelope are considered as
true cluster members. In this way 3334, 943 and 499 member stars
are found in NGC 6067, NGC 2506 and IC 4651 respectively.
 
\begin{figure}
    \begin{center}
    \centering
    \includegraphics[width=9cm,height=9cm]{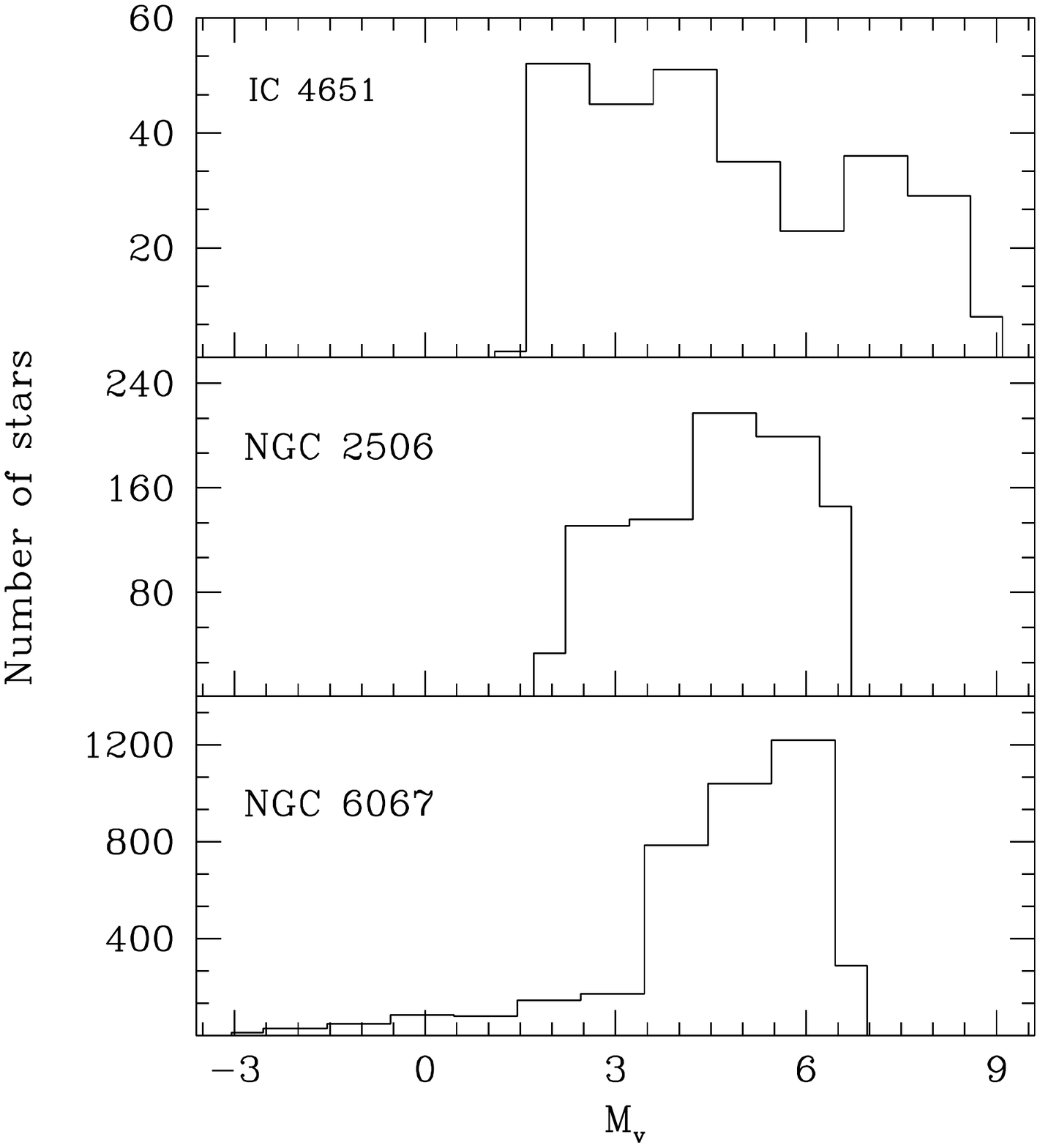}
    \caption{Histograms for the most probable main sequence stars in absolute
    magnitude bin for the clusters under study.}
    \label{lf}
    \end{center}
  \end{figure}

To construct luminosity function we first converted the apparent $V$ magnitude into
absolute one, using the distance and reddening calculated in section \ref{sec:parallax} and
\ref{sec:iso} respectively. Now we calculated the number of stars in each magnitude bin. 
After correcting the data for incompleteness we plotted
luminosity function for the clusters NGC 6067, 
NGC 2506 and IC 4651 which are shown in Fig.\ref{lf}.

Fig. \ref{lf} shows that the luminosity function of NGC 6067 increases towards the fainter
magnitude. But after $M_{v} \sim 6 $ mag, we observed a dip in the
luminosity function of the cluster. This cluster is young 
and hence faint stars are still bound with the cluster.
The luminosity function of NGC 2506 seems very flat except one 
prominent peak observed at $M_{v} \sim $ 4.5 mag. The flatness
signature may occur due to the
dynamical evolution of the cluster, which causes the evaporation of the 
low mass stars from the cluster. Similar results were also reported by
\citet{2012MNRAS.425.1567L} for the clusters NGC 1245 and NGC 2506.
The luminosity function of IC 4651 decreases towards the fainter
magnitude. From the histogram, we can conclude that 
most of the low mass stars have evaporated from the cluster 
due to the dynamical evolution. 

%mf
%---------------------------------------------------------------------------

\subsection{Mass function} \label{subsec:m}

The mass function (MF) denotes the relative number of stars in a unit mass range.
The luminosity function of a cluster can be converted into mass-function using 
theoretical evolutionary tracks.
For this conversion, we used theoretical models given by \citet{2017ApJ...835...77M}.
The slope of mass function
is calculated, using the relation   \\

$log \frac{dN}{dM} = -(1+x) \times log(M) + constant$   \\

\begin{figure}
    \centering
    \includegraphics[width=8cm]{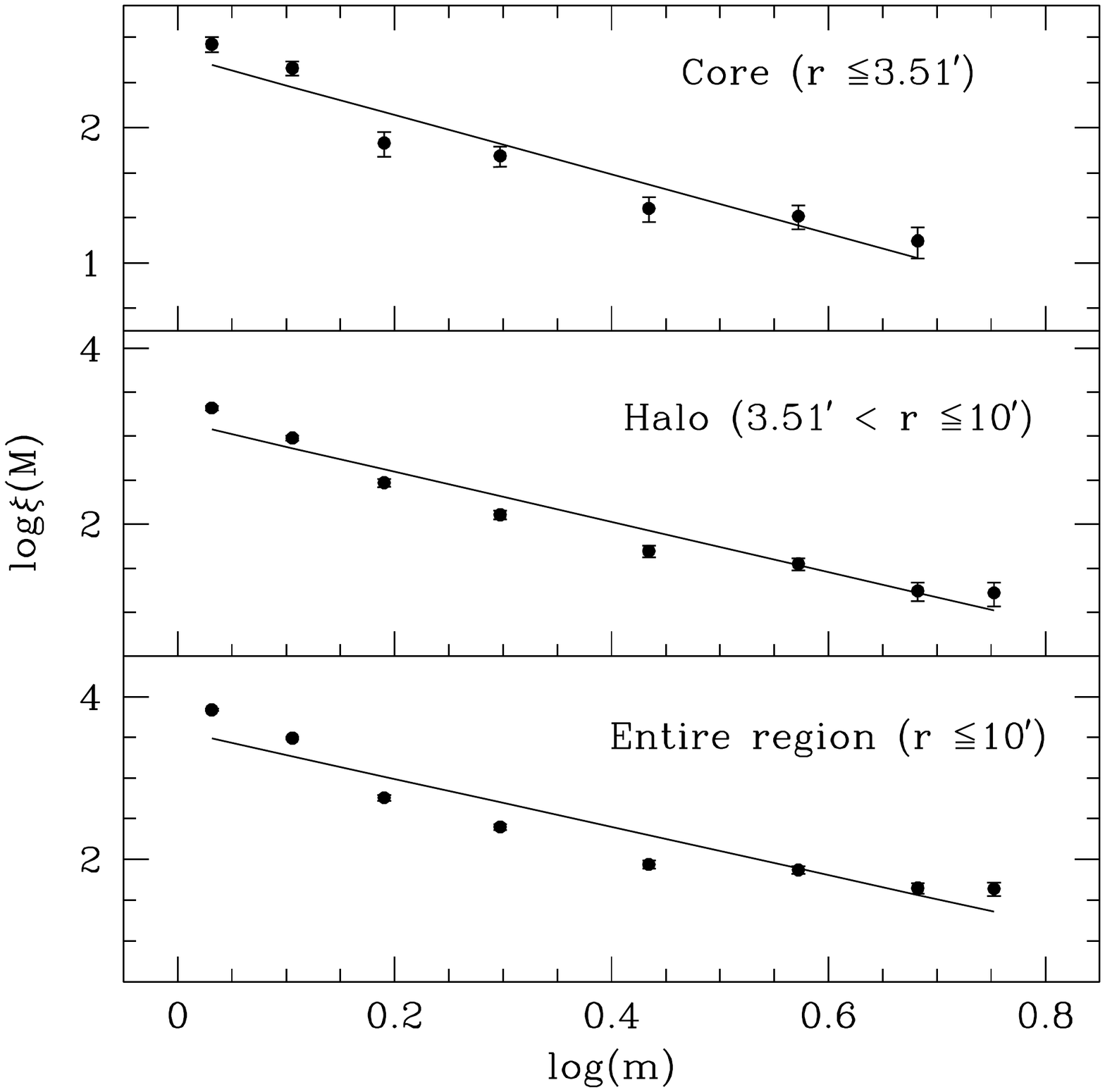}
    \caption{The plots of the mass-function for NGC 6067 in core,
     halo and entire region of the cluster. Filled circles shows data points
    and bars shows their standard error. Straight lines represent
     the least square fitting to the data points {\bf and provide the value of $(1+x)$.}}
    \label{mf6067}
  \end{figure}

\begin{figure}
    \centering
    \includegraphics[width=8cm]{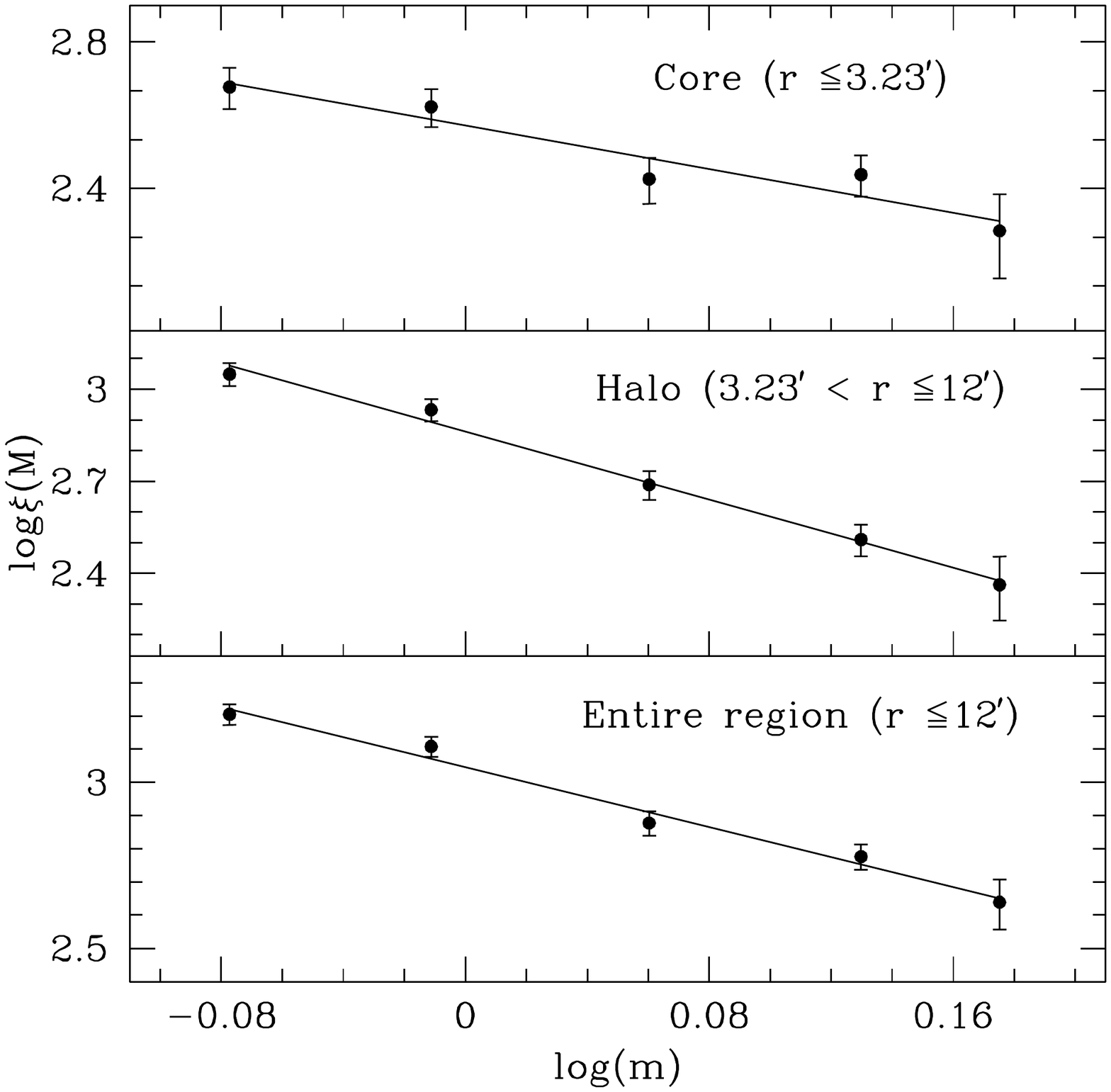}
    \caption{Same as Fig. \ref{mf6067} for the cluster NGC 2506}
    \label{mf2506}
  \end{figure}

\begin{figure}
    \centering
    \includegraphics[width=8cm]{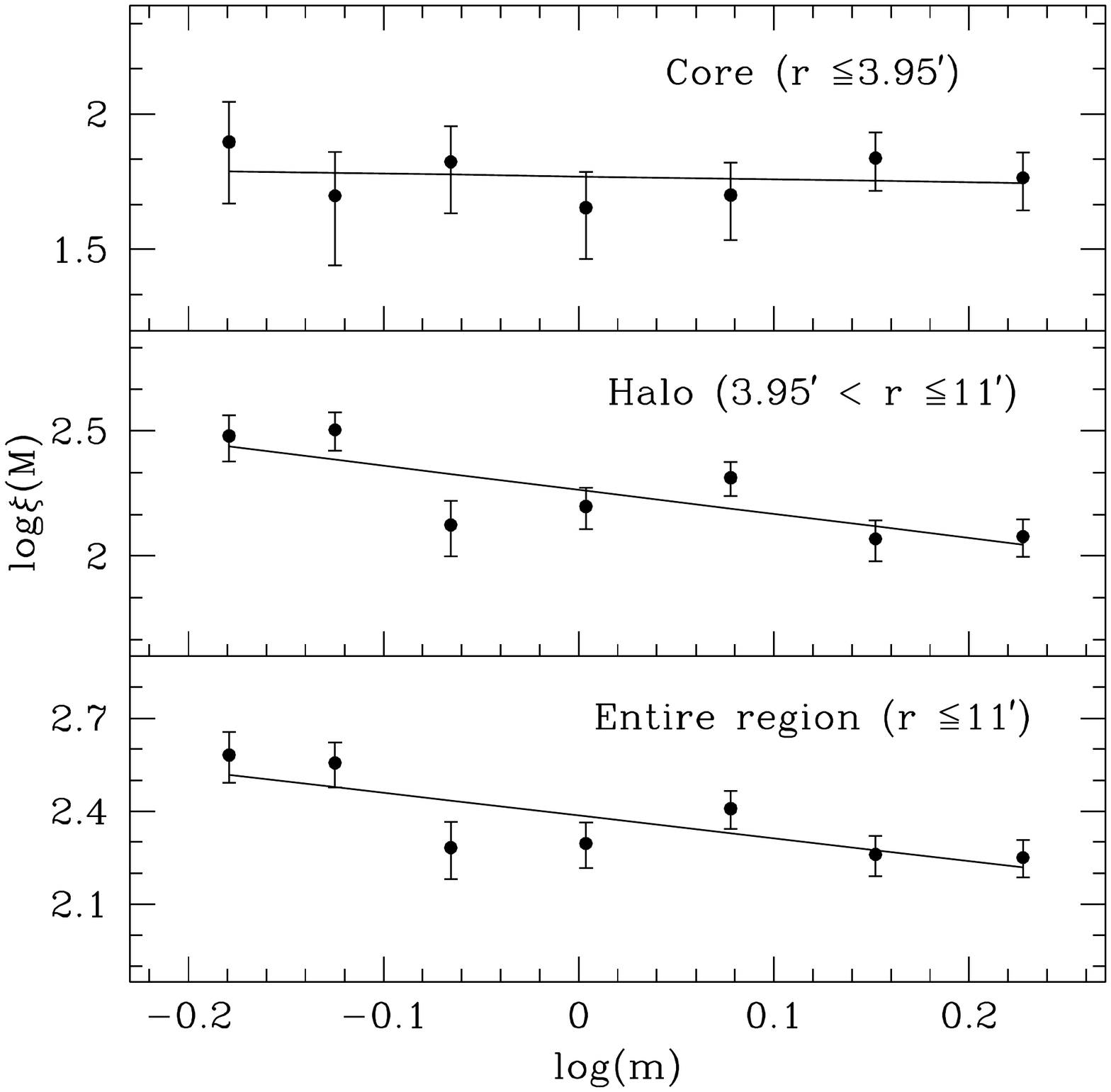}
    \caption{Same as Fig. \ref{mf6067} for the cluster IC 4651}
    \label{mf4651}
  \end{figure}

Where $dN$ is the number of stars in mass bin $dM$ and $M$ is the central mass 
of the bin. From this equation, we can determine the value of 
mass function slope $x$. In Fig. \ref{mf6067}, \ref{mf2506} and
\ref{mf4651} the mass distribution $\xi(M)$
is shown for the clusters NGC 6067, NGC 2506 and IC 4651 respectively.
The mass function is calculated for three
regions i.e, core, halo and the entire region. The radius of the cluster derived 
in section \ref{sec:cen} is taken for entire region while core and halo region are 
considered as core radius and beyond the core region respectively. 
The mass function slopes derived by least-square fitting for all regions 
are listed in Table \ref{mftab} for all the clusters.

\begin{table}
   \tiny
   \centering
   \caption{ The mass-function slopes for the core,
     halo and entire regions of the clusters under study.
   }
   \begin{tabular}{lcrcr}
   \hline\hline
  Cluster & Mass  &      \multicolumn{3}{c} {Mass function slope ($x$)}   \\
  %\cline{3-5} 
        & range $(M_{\odot})$    & Core  & Halo  & Entire region  \\
  \hline
  NGC 6067 &  1.00 - 5.99 & $1.19 \pm 0.30$ & $bf 1.85 \pm 0.28$ & $ 1.96 \pm 0.42 $  \\
  NGC 2506 & 0.77 - 1.54  & $ 0.49 \pm 0.26$ & $ 1.77 \pm 0.14$ &  $ 1.26 \pm 0.16$  \\
  IC 4651  & 0.62 - 1.83  & $ -0.89 \pm 0.26$ & $ -0.04 \pm 0.35$ & $ -0.26 \pm 0.27 $ \\
  \hline
  \end{tabular}
  \label{mftab}
  \end{table}
 
For NGC 6067 the mass-function slope of the entire cluster and halo region
is greater than Salpeter value $x = 1.35$ \citep{1955ApJ...121..161S}. 
Mass function slope for core region of NGC 6067 and 
entire cluster region of NGC 2506 are found
comparable with the Salpeter value.
A flatter mass-function slope is observed for IC 4651 in all regions. 
Table \ref{mftab} also shows that mass-function slope becomes flat when one 
goes from halo to core regions in all the clusters. 
This may be due to the mass segregation in the clusters.
A similar spatial variation in mass function slope was also observed by 
\citet{1992BASI...20..287P}
for nine clusters and \citet{2001A&A...375..840D} 
for the clusters king 7 and Be 20. 

%mseg
%------------------------------------------------------------------------

\subsection{Mass segregation} \label{sec:msegg}

\begin{figure}
    \centering
    \includegraphics[width=9cm,height=9cm]{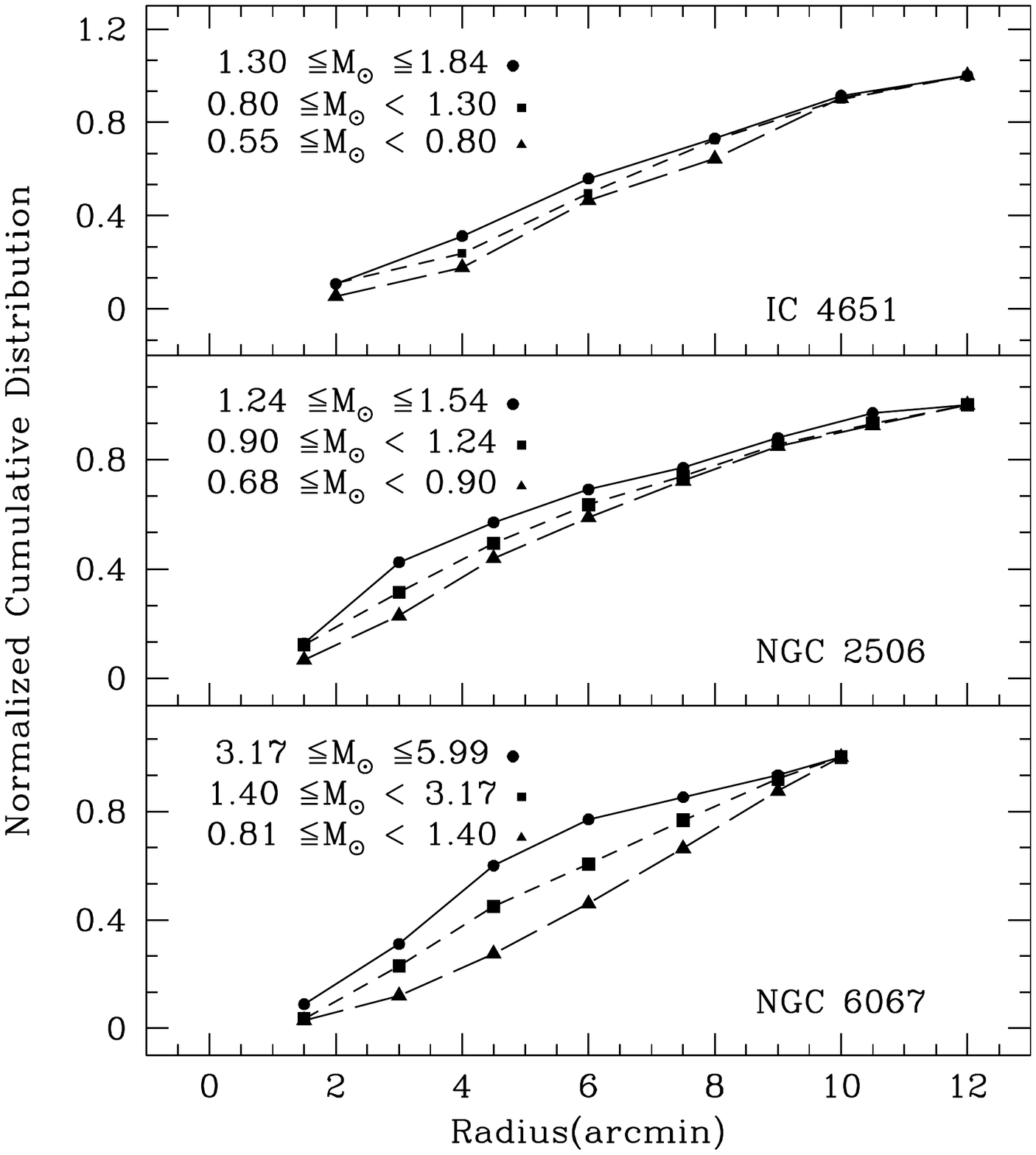}
    \caption{The figure shows normalized cumulative distribution
    of stars in cluster regions. The inner most curve is for the
    massive stars and the outer most curve is for lighter one.
    }
    \label{ms}
  \end{figure}

To study the signature of 
mass segregation in the clusters under study, we 
divided the stars into three mass range as shown in Fig. \ref{ms}.
Cumulative radial stellar distributions for different regions 
are shown in Fig. \ref{ms}.
 
Fig. \ref{ms} shows that for the clusters NGC 6067 and IC 4651,
high mass stars are located towards the cluster centre while
low mass stars are distributed in the outer region of the cluster.
This indicates a clear mass segregation effect in these clusters.
On the other hand, NGC 2506 shows less mass segregation in comparison
to NGC 6067 and IC 4651.
To check whether these curves belong to the same distribution or not, we 
performed the Kolmogorov-Smirnov (K-S) test for different mass range.
From K-S test, we found 99, 90 and 70 $\%$  confidence level
for the clusters NGC 6067, NGC 2506 and IC 4651 respectively.

In open star clusters, dynamical evolution may be a possible reason for
the presence of mass-segregation.
In the very beginning, a cluster may possess a random spatial stellar mass distribution
and due to the dynamical evolution, this mass distribution gets modified.
The dynamical evolution takes place in a time scale over which
individual cluster member exchange energy and their velocity distribution
approach a Maxwellian equilibrium. 
It is also possible that
mass segregation may be present from the time of the birth of the cluster.
To check these two possibilities for the presence of mass-segregation,
we calculated
the dynamical evolution time for clusters using
the relation \\

$T_{E} = \frac{8.9 \times 10^{5} N^{1/2} R_{h}^{3/2}}{\langle m \rangle^{1/2} log(0.4N)}$     \\

where $N$ is the number of cluster members, $R_{h}$ is the radius containing
half of the cluster members. For this analysis, we used half of the 
cluster radius as $R_{h}$, $\langle m \rangle$ is mean cluster mass 
\citep{1971ApJ...164..399S}.
We used most probable cluster members which are selected using
both vector point diagrams and photometric criteria. 
%We also applied the completeness corrections on these stars.

The dynamical evolution time is determined as 396, 429 and 294 Myr
for the clusters NGC 6067, NGC 2506 and IC 4651 respectively.
On comparing these values with the cluster age, we find that
NGC 6067 does not show relaxation whereas NGC 2506 and
IC 4651 achieved the Maxwellian equilibrium. It concludes that
mass-segregation observed in NGC 6067 may be due to the imprint of star formation
and in NGC 2506 and IC 4651 it may be due to the dynamical
evolution or imprint of cluster formation or both.

%conclusion
%-------------------------------------------------------------------------

\section{Conclusions} \label{con}

In this article, a combination of high precision Gaia DR2 proper motion data
and ground-based wide-field photometric data are used to probe the dynamical status
and orbits of three open star clusters NGC 6067, NGC 2506 and IC 4651. The present
analysis is based on the cluster members selected by using proper motion data
and parallax data.
The main results of the present analysis are the following: \\

\begin{enumerate}

\item We used Gaia DR2 proper motion and parallax data to separate cluster
members from field stars and also calculated mean
proper motions for the clusters. These are found as
$1.90 \pm 0.01$ and $-2.57 \pm 0.01$ mas$yr^{-1}$, $-2.57 \pm 0.01$ and $3.92 \pm 0.01$ mas$yr^{-1}$,
$-2.41 \pm 0.01$ and $-5.05 \pm 0.02$ mas$yr^{-1}$ in RA and DEC directions
for the clusters
NGC 6067, NGC 2506 and IC 4651 respectively.

\item We determined average parallax and corresponding heliocentric distances 
as $3.01 \pm 0.87$ kpc, $3.88 \pm 0.42$ kpc and $1.00 \pm 0.83$ kpc
for the clusters
NGC 6067, NGC 2506 and IC 4651 respectively.

\item Galactic orbits and orbital parameters are determined for the
clusters, using Galactic potential models. We found that all three
clusters are orbiting in a boxy orbit. NGC 6067 and IC 4651 have small
perigalactic and apogalactic distances as compared to NGC 2506.

\item From radial density profiles of clusters, 
radii are found as $10^{\prime}$, $12^{\prime}$
and $11^{\prime}$ for the clusters NGC 6067, NGC 2506 and IC 4651 respectively.
The corresponding linear sizes are 6.81, 12.43 and 3.03 pc.

\item Ages of 
the clusters are found as $66 \pm 8$ Myr, $2.09 \pm 0.14$ Gyr and $1.59 \pm 0.14$ Gyr for 
the clusters NGC 6067, NGC 2506 and IC 4651 respectively.

\item The luminosity function of NGC 6067 increases upto $M_v$ $\sim 6$ and decreases towards
the fainter end. This implies that fainter stars are still bound in NGC 6067. 
A flat luminosity function is observed
for NGC 2506 while a decreasing luminosity function is found for IC 4651.
Due to dynamical evolution, fainter stars
of NGC 2506 and IC 4651 may get evaporated from the cluster region.

\item Mass function slope for core region of NGC 6067 and
entire cluster region of NGC 2506 is found comparable
with the \citet{1955ApJ...121..161S} value whereas for IC 4651 
it is flatter.
We have also found that for all the clusters under study,
the slope becomes flattered towards the 
core of the clusters. We found a hint of 
mass-segregation in all the three cluster in our sample.

%\item From present analysis it can be concluded that NGC 2506 is the oldest cluster
%in the sample even though it's stars are bound to the cluster. IC 4651 is slightly
%younger than NGC 2506 and it is a sparse cluster. The reason behind survival of
%NGC 2506 is may be that the
%location of its birth place is far from Galactic disc and also
%it is orbiting in outer disc.
%Due to this, NGC 2506 is not affected
%by the Galactic tides as compared to the other two cluster.

\end{enumerate}

\section*{Acknowledgements}

We are thankful to the anonymous referee for careful reading 
and constructive suggestions that improved the 
overall quality of the paper.
This research is based on data obtained from ESO Science Archive
Facility.
This work has made use of data from the European Space Agency (ESA) mission
{\it Gaia} (\url{https://www.cosmos.esa.int/gaia}), processed by the {\it Gaia}
Data Processing and Analysis Consortium (DPAC,
\url{https://www.cosmos.esa.int/web/gaia/dpac/consortium}). Funding for the DPAC
has been provided by national institutions, in particular the institutions
participating in the {\it Gaia} Multilateral Agreement.
This study also made use of WEBDA database. This work
is partially supported by the Natural Science Foundation of China
(NSFC-11590782, NSFC-11421303).

%%%%%%%%%%%%%%%%%%%% REFERENCES %%%%%%%%%%%%%%%%%%

\bibliographystyle{mnras}
\bibliography{article_geeta_final}

%%%%%%%%%%%%%%%%%%%%%%%%%%%%%%%%%%%%%%%%%%%%%%%%%%

\label{lastpage}
\end{document}